\documentclass[aps,preprint,superscriptaddress,floatfix,showpacs,amsmath,amssymb,nofootinbib,longbibliography,nobibnotes,noeprint]{revtex4-2}

\usepackage{graphicx}
\usepackage{dcolumn}
\usepackage{bm}
\usepackage{color}
\usepackage{tabularx}
\usepackage{csquotes}
\usepackage[normalem]{ulem}

\usepackage{amsmath}
\usepackage{nicefrac}
\usepackage{booktabs, multirow}
\usepackage{mathtools}
\usepackage{SIunits}

\newcommand{\cpp}{$c_{\rm p}$}
\newcommand{\cpm}{$c_{\rm p}^{\rm m}$}
\newcommand{\mb}{\(\mu _{\rm B}\)}

\newcommand{\tn}{$T_{\rm N}$}

\newcommand{\lfs}{Li\(_2\)FeSiO\(_4\)}
\newcommand{\glfs}{$\gamma_{\rm II}$-Li\(_2\)FeSiO\(_4\)}

\newcommand{\lzs}{Li\(_2\)ZnSiO\(_4\)}

\newcommand{\jmk}{J/(mol$\cdot$K)}

\newcommand{\Bhyp}{$B_{\rm{hyp}}$}

\begin{document}

\title{Quasi-two-dimensional magnetism and antiferromagnetic ground state in Li$_2$FeSiO$_4$}


\author{W.~Hergett}
\affiliation{Kirchhoff Institute of Physics, Heidelberg University, INF 227, D-69120 Heidelberg, Germany}

\author{N.~Bouldi}
\affiliation{Institute for Theoretical Physics, D-69120 Heidelberg, Germany}

\author{M.~Jonak}
\affiliation{Kirchhoff Institute of Physics, Heidelberg University, INF 227, D-69120 Heidelberg, Germany}

\author{C.~Neef}
\affiliation{Kirchhoff Institute of Physics, Heidelberg University, INF 227, D-69120 Heidelberg, Germany}

\author{C.~Ritter}
\affiliation{Institut Laue-Langevin, 38042 Grenoble, France}

\author{M.~Abdel-Hafiez}
\affiliation{Kirchhoff Institute of Physics, Heidelberg University, INF 227, D-69120 Heidelberg, Germany}
\affiliation{Physics Department, Faculty of Science, Fayoum University, Fayoum 63514, Egypt}
\affiliation{Department of Applied Physics and Astronomy, University of Sharjah, P. O. Box 27272 Sharjah, United Arab Emirates}

\author{F. Seewald}
\affiliation{Institute for Solid State and Materials Physics, TU Dresden, D-01069 Dresden, Germany}

\author{H.-H. Klauss}
\affiliation{Institute for Solid State and Materials Physics, TU Dresden, D-01069 Dresden, Germany}

\author{M.~W.-Haverkort}
\email[Email:]{haverkort@itp.uni-heidelberg.de}
\affiliation{Institute for Theoretical Physics, D-69120 Heidelberg, Germany}

\author{R.~Klingeler}
\email[Email:]{klingeler@kip.uni-heidelberg.de}
\affiliation{Kirchhoff Institute of Physics, Heidelberg University, INF 227, D-69120 Heidelberg, Germany}


\date{\today}

\begin{abstract}

Our experimental (neutron diffraction, Mössbauer spectroscopy, magnetic susceptibility, specific heat) and numerical studies on the evolution of short- and long-range magnetic order in $\gamma_{\rm II}$-Li\(_2\)FeSiO\(_4\) suggest a quasi-two-dimensional (2D) nature of magnetism. The experimental data obtained on single crystals imply long-range antiferromagnetic order below $T_{\rm N}= 17$~K. A broad maximum in magnetic susceptibility $\chi$ at $T_{\rm m}\simeq 28$~K, observation of magnetic entropy changes up to 100~K and anisotropy in $\chi$ are indicative of low-dimensional magnetism and suggest short-range magnetic correlations up to 200~K. Neutron diffraction shows that long-range antiferromagnetic order is characterised by the propagation vector \textbf{k}=(\nicefrac{1}{2},0,\nicefrac{1}{2}). The ordered moment $\mu = 2.50(2)$~\mb /Fe, at $T = 1.5$~K, is along the crystallographic $a$-axis. This is consistent with the observed static  hyperfine field of $B_{\rm hyp}=14.8(3)$\,T by Mössbauer spectroscopy which indicates significant orbital contributions. The temperature dependence of $B_{\rm hyp}$ yields the critical exponent $\beta=0.116(12)$ which is in the regime of the 2D Ising behaviour. LSDA+U studies exploiting the experimental spin structure suggest dominating magnetic exchange coupling within the $ac$-layers (i.e., $J_3\simeq -6$~K and $J_6\simeq-2$~K) while interlayer coupling is much smaller and partly frustrated. This confirms the 2D nature of magnetism and is in full agreement with the experimental findings.
\end{abstract}

\maketitle

\section{Introduction}

\begin{figure}[htb]
	\includegraphics[width=1.0\columnwidth,clip] {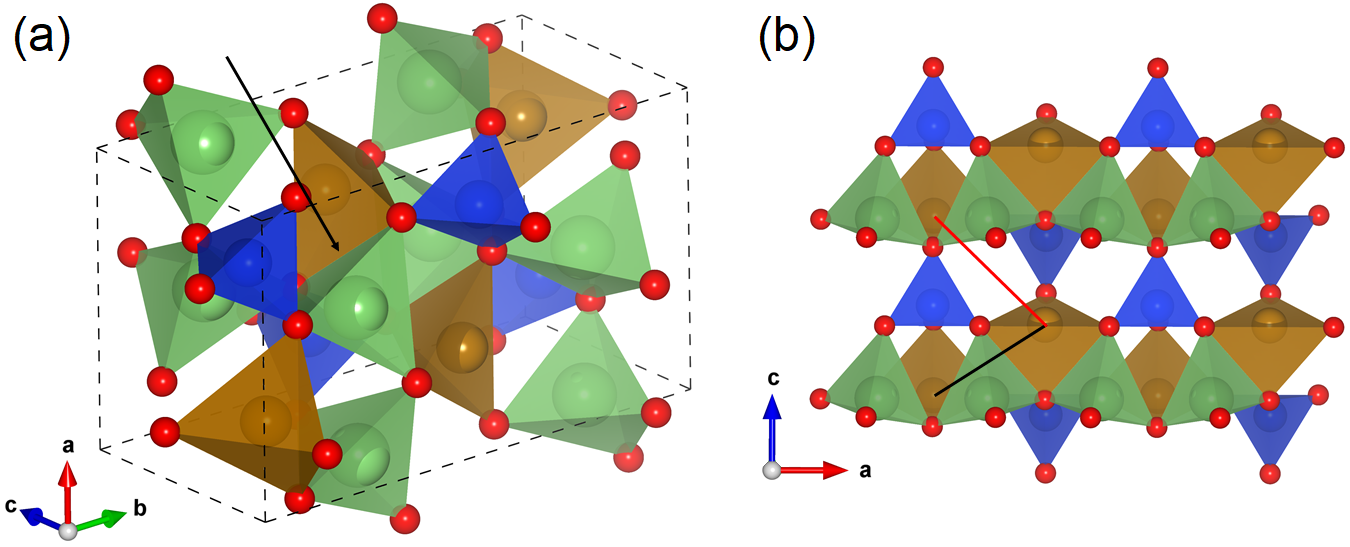}
	\caption{(a) Crystallographic unit cell of $\gamma_{\rm II}$-\lfs, consisting of tetrahedra of LiO\(_4\) (green), FeO\(_4\) (brown), and SiO\(_4\) (blue). Oxygen ions are depicted in red. The black arrow points to the edge shared by the FeO\(_4\) and LiO\(_4\) tetrahedra. (b) Cross-section of the $ac$-plane. The black and red lines mark the shortest and  second-shortest magnetic bonds, respectively.}\label{struc}
\end{figure}

The occurrence of magnetic order in low-dimensional and/or magnetically frustrated spin systems is directly linked to the interplay of magnetic interaction, geometric arrangement of the magnetic centres, and magnetic anisotropy. While in pure two-dimensional (2D) Heisenberg systems no long-range order is expected at finite temperature, observation of long-range magnetic order down to the monolayer in transition-metal based van-der-Waals materials highlights the relevance of magnetic anisotropy in the evolution of a magnetic ground state (see, e.g., Ref.~\onlinecite{Gong2019review} and references therein). In particular in Jahn-Teller active systems the orbital degree of freedom may be relevant, too, so that for example low-dimensional magnetism may be realised as observed in a variety of transition metal oxides~\cite{Imada1998,Tokura2000,Oles2010}. Spin-orbit entanglement can yield magnetically and orbitally ordered ground states as observed, e.g., paradigmatically in systems such as KCuF$_3$ or LaMnO$_3$ and its doped variants~\cite{Caciuffo2002,Feiner,Geck2005}. 
The orthosilicate \lfs\ has been  intensively studied in polycristalline form as a high-capacity cathode material for lithium-ion batteries~\cite{Islam2011,Sirisopanaporn2011,Yang2016}. Its orthorhombic $Pmnb$-structured polymorph, \glfs, whose crystallographic unit cell is sketched in Fig.~\ref{struc}a, exhibits tetrahedrally coordinated Fe$^{\rm 2+}$-ions arranged in a layered structure where alternating layers of FeO$_4$-/SiO$_4$- and LiO$_4$-tetrahedra are stacked along the $b$-axis.~\cite{Lu2015,Nishimura2008,Boulineau2010} The distance between adjacent Fe$^{2+}$-ions is 4.115(2)~\AA\ (black line in Fig.~\ref{struc}b), while the next-nearest neighbors are separated by 4.683(2)~\AA\ (red line in Fig.~\ref{struc}b). Incidentally, both -- the nearest and next-nearest -- neighbors lie almost perfectly in the $ac$-plane, with only a small displacement along the $b$-axis. While, Fe$^{\rm 2+}$-ions are well-separated along the $b$-axis showing a nearest-neighbor distance of 5.339(3)~\AA . In tetrahedral coordination, 3$d$-orbitals split into lower-lying $e_{\rm g}$- and higher-lying $t_{\rm 2g}$-orbitals. The high-spin $S = 2$ configuration $e_{\rm g}^3t_{\rm 2g}^3$ of Fe$^{\rm 2+}$ ions is Jahn-Teller (JT) active and implies the relevance of orbital degrees of freedom. 

Here, we report the evolution of short- and long-range magnetic order in $\gamma_{\rm II}$-Li\(_2\)FeSiO\(_4\) single crystals and determine the magnetic ground state, which is characterised by the propagation vector \textbf{k}=(\nicefrac{1}{2},0,\nicefrac{1}{2}). Our numerical studies, based on the experimental spin structure, imply dominating magnetic exchange coupling within the $ac$-layers while interlayer coupling is small and partly frustrated. The numerically suggested quasi-two-dimensional nature of magnetism in \lfs\ is confirmed by our observation of a broad correlation-type maximum at $T_{\rm m}\simeq 28$~K and of short-range magnetic correlations more than ten-times above the long-range antiferromagnetic  ordering temperature $T_{\rm N}= 17$~K.

\section{Experimental and numerical methods}

Millimeter-sized single crystals of \lfs\ were grown by the high-pressure optical floating-zone method as described in detail in Refs.~\onlinecite{Hergett2019a,Hergett2019b,Hergett2021}. The thus-grown single crystals were oriented and cut into cuboids with approximate dimensions $1.1 \times 1.1 \times 1.0$~mm$^3$. Magnetization was studied in a Quantum Design MPMS-XL5 SQUID magnetometer. Specific heat measurements were performed in a Quantum Design Physical properties Measurement system (PPMS-14) using the relaxation method. 

Polycrystalline samples of $Pmnb$-\lfs\ were synthesized by solid-state reaction as described in Ref.~\onlinecite{Hergett2019b}. Phase purity was confirmed by X-ray diffraction (XRD), magnetic susceptibility, and powder neutron diffraction (PND). \lzs\ was synthesized as a non-magnetic analog to \lfs\ through a conventional solid-state reaction route, too. Stoichiometric amounts of Li\(_2\)CO\(_3\) ZnO, and SiO\(_2\) were mixed, ball-milled, calcined at 1100~°C, and characterized for phase purity by XRD. Neutron diffraction measurements on the powder sample of $Pmnb$-\lfs\ were obtained at several temperatures on the high intensity D20 powder diffractometer at the Institute Laue Langevin, Grenoble, France.~\cite{ill} The sample, with a mass of approximately 1.6~g, was placed in a sealed vanadium can within a He cryostat. For both magnetic structure determination and nuclear structure refinement an incident neutron wavelength of 2.41\,\AA\ was used. Rietveld refinements of PND data were performed using the FullProf Suite program~\cite{Fullprof1993} and the magnetic symmetry analysis was performed by means of the program BasIreps~\cite{Basireps}, included in the FullProf Suite package. Sketches of the crystallographic and magnetic structures were drawn with the program VESTA~\cite{Vesta}.

$^{57}$Fe-Mössbauer (MB) measurements were performed on a powder sample from the same batch as studied by neutron diffraction and magnetisation. The sample was mounted in a CryoVac helium flow cryostat using a commercial WissEL Mössbauer spectrometer. A $^{57}$Co in Rh source driven in sinusoidal mode was used. Data evaluation was performed using the Mössfit software package~\cite{HyInt} using a transmission integral analysis to account for the sample thickness. Measurements were carried out at temperatures ranging from 4.2~K to 296~K. All isomer shifts are reported relative to $\alpha$-Fe at room temperature.

For the numerical studies, scalar relativistic first-principles calculations based on density functional theory (DFT) were performed by means of the package FPLO\cite{Koepernik_Full_potential_1999,Opahle1999} (Full-Potential Local-Orbital) using the local density approximation (LDA) with the Perdew-Wang~92 parametrization\cite{Perdew_accurate_1992} for the exchange correlation functional and periodic boundary conditions. FPLO is a code to solve the Kohn-Sham equation with a basis of atomic-like local orbitals within a full-potential approach. The experimental crystallographic structure (see Table \ref{tab:criststructure}) was used for the calculations, and a $\Gamma$-centered \textit{k}-point grid  $8 \times 5 \times 10$ was set for the unit cell (4 Fe atoms).

\begin{table}[tb]
\caption{Lattice parameters and atomic coordinates in the conventional unit cell for \glfs\ used in the calculation (determined and refined by single-crystal X-ray diffraction).~\cite{Hergett2019b}}
\label{tab:criststructure}
\begin{tabular}{@{\extracolsep{4pt}}lcccc}
\hline
\hline
\multicolumn{5}{l}{Li$_2$FeSiO$_4$, space group $Pmnb$} \\  [3pt]
\hline
\multicolumn{5}{l}{$a$ = 6.27837~$\angstrom$; $b$ = 10.6290~$\angstrom$; $c$ = 5.03099~$\angstrom$} \\
\hline
 &  & $x$ & $y$ & $z$ \\

Li (8\textit{d}) & 1 & 0.494739 & 0.331319 & -0.707234 \\
Fe (4\textit{c}) & .m. & 0.750000 & 0.418512 & -0.299501 \\
Si (4\textit{c}) & .m. & 0.750000 & 0.583922 & 0.193420 \\
O (4\textit{c}) & .m. & 0.750000 & 0.436406 & -0.716996 \\
O (4\textit{c}) & .m. & 0.750000 & 0.590409 & -0.131224 \\
O (8\textit{d}) & 1 & 0.461921 & 0.343885 & -0.305747 \\
\hline
\hline
\end{tabular}
\end{table}

To obtain the total energies associated with several magnetic configurations, we used the magnetic generalization of LDA with inclusion in the Fe 3-$d$ shell of the Hubbard term  (LSDA+U)\cite{Eschrig_density_2003} with values for the Slater integrals $F^0$~=~6.5 eV, $F^2$~=~10.1 eV and $F^4$~=~6.2 eV. To simulate antiferromagnetic structures, several supercells containing 8 iron atoms were built. We used the supercells  $1 \times 1 \times 2$ and $2 \times 1 \times 1$ as well as a more unusual supercell of lattice parameters $\mathbf{a}'=2\mathbf{a}$, $\mathbf{b}'=\mathbf{b}$ and $\mathbf{c}'=\mathbf{c}-\mathbf{a}$. The latter was needed to model the magnetic state obtained by neutron diffraction. In each case, the \textit{k}-point grid was reduced according to the length multiplication.

\section{Experimental Results}

\subsection{Macroscopic properties of single crystals}\label{TD}

\begin{figure}[b]
\includegraphics[width=1.0\columnwidth,clip]{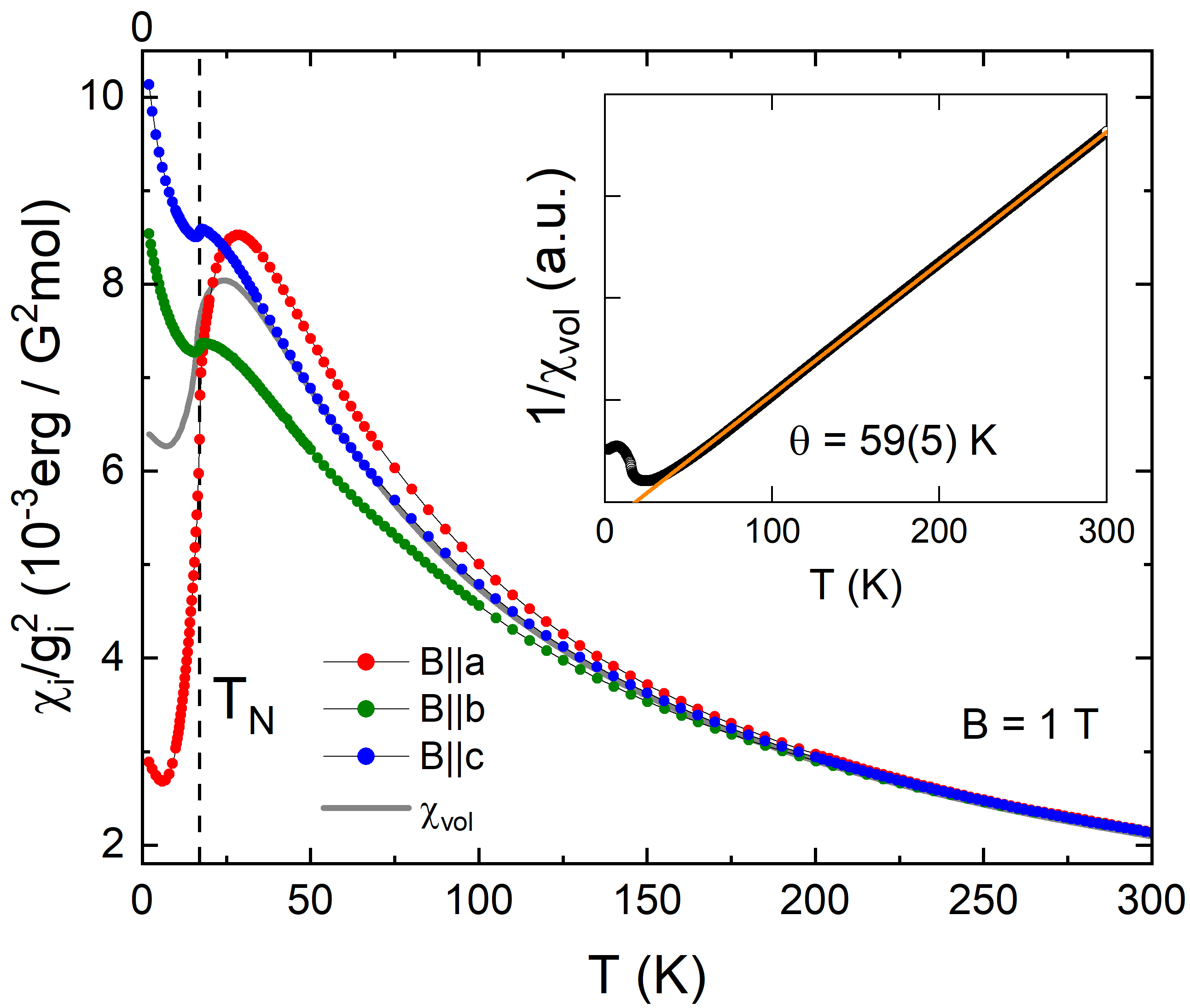}
\caption{Static magnetic susceptibility, $\chi_{\rm i} = M_{\rm i}/B_{\rm i}$ ($i=a,b,c,{\rm vol}$), obtained at $B=1$~T, divided by axis-dependent $g$-factors, $g_{\rm i}$, as a function of temperature. The dashed vertical line marks \tn. Inset: Inverse of the volume susceptibility, $\chi_{\rm vol}$, vs. temperature at $B=5$~T. The orange line represents a fit using a Curie-Weiss model (see the text).}
\label{chi}
\end{figure}

Static magnetic susceptibility $\chi = M/B$ of \lfs\ exhibits anisotropic behaviour up to the highest measured temperatures 
This high-temperature anisotropy in $\chi$ is associated with the $g$-tensor, as shown by scaling $\chi$ with axis-dependent $g$-factors in Fig.~\ref{chi}. At high temperatures, $\chi$ reveals Curie-Weiss-like behaviour, which is confirmed by quasi-linear temperature dependence of the inverse volume susceptibility $\chi^{-1}_{\rm vol} = 3/(\chi_{\rm a}+\chi_{\rm b}+\chi_{\rm c})$ at $T\gtrsim 200$~K (see the inset of Fig.~\ref{chi}). Analyzing the volume susceptibility in terms of a Curie-Weiss-like model, $\chi = N_{\rm A}p^2\mu_{\rm B}^2/3k_{\rm B}(T+\Theta)+\chi_0$, where $N_{\rm A}$  is Avogadro's number, \mb\ is the Bohr magneton, and $k_{\rm B}$ is Boltzmann's constant, yields a Weiss temperature of $\Theta\approx 59(5)$~K and an  effective magnetic moment of $p=5.57(10)$~\mb /f.u.. The sign of $\Theta$ indicates predominant antiferromagnetic interactions. From the effective magnetic moment, the tetrahedrally coordinated Fe$^{2+}$ ions are deduced to be in the high-spin $S=2$-state, suggesting an electronic configuration of $3e_{\rm g}^3t_{\rm 2g}^3$. We conclude the volume $g$-factor $g_{\rm vol}=2.27(1)$ which is in the upper range of typical values of tetrahedrally coordinated Fe$^{\rm 2+}$.~\cite{Krzystek,Gibart,Fritsch} From the uniaxial susceptibilities we read-off $g_{\rm a}=2.36(1)$, $g_{\rm b}=2.19(1)$, and $g_{\rm c}=2.28(1)$.

Upon cooling, magnetic anisotropy beyond the $g$-tensor anisotropy appears below $T \approx 200$~K.  A sharp decrease of $\chi$ at low temperatures implies the onset of long-range antiferromagnetic order with the crystallographic $a$-direction being the magnetic easy axis. In addition, for $B\|a$ there is a broad maximum in $\chi_{\rm a}$ at around $T_{\rm m} = 28$\,K while the onset of long-range antiferromagnetic order is signaled by a sharp downturn in $\chi_{\rm a}$ and a $\lambda$-like anomaly in the magnetic specific heat, $\partial (\chi_{\rm a}T)/\partial T$ (see Figs.~\ref{chi} and \ref{sm2}). These data imply $T_{\rm N} = 17.0(5)$\,K which is similar to previous studies on polycrystalline \lfs ~\cite{Gong2011,Bini2013,Zaghib2006}. For $B\perp a$, the magnetic susceptibility displays much smaller anomalies at \tn . A Curie-like upturn at lowest temperatures indicates the presence of approximately 0.5~\% of only weakly correlated magnetic moments that do not participate in long-range antiferromagnetic spin order and may be considered 'quasi-free'. The presence of quasi-free moments is also confirmed by the initial Brillouin-like right-bending observed in the $M$ vs. $B$ curves 
A small fraction of quasi-free Fe$^{2+}$-moments is expected due to defects which may in particular be Li-Fe antisite defects which arise from similar covalent radii of Li and Fe ions~\cite{Hergett2019b,Werner2020,Neef2017}. These antisite defects typically exhibit an anisotropic nature, as reflected by the fact that signatures of quasi-free moments in $M(T)$ and $M(B)$ are much weaker for $B||a$ axis. Our data indicate an $g$ factor anisotropy of the quasi-free moments of $g^{\rm qf}_b\simeq g^{\rm qf}_c \simeq 4g^{\rm qf}_a$.

\begin{figure}[htb]
\includegraphics[width=1.0\columnwidth,clip]{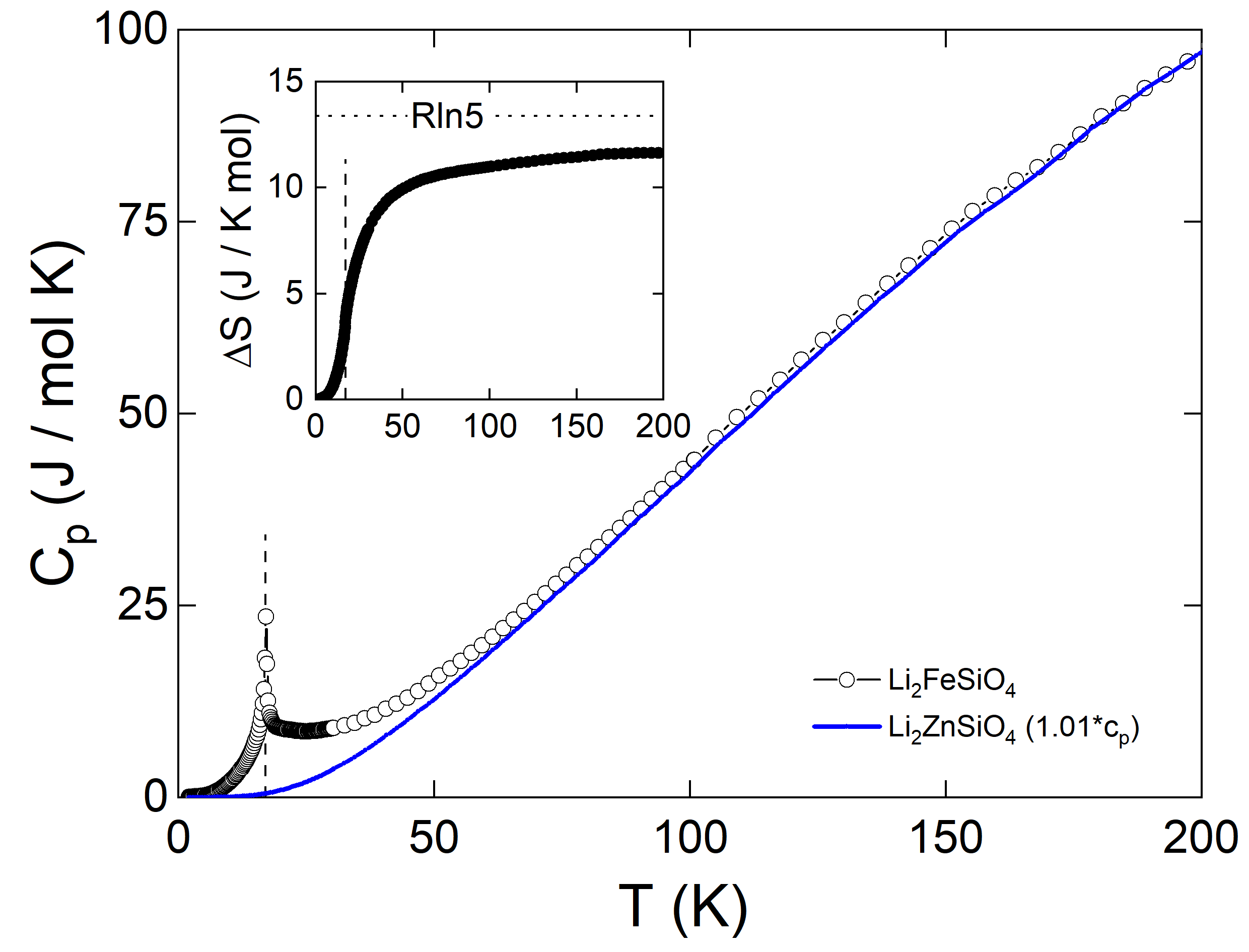}
\caption{Specific heat capacity, \(c_p\), of \lfs\ (black symbols) and \lzs\ (blue line, scaled by a factor of 1.01) at \(B=0\)~T vs. temperature. (a) \(c_p\) vs. \(T^3\) 
Inset: Magnetic entropy changes, \(\Delta S\), obtained by integrating \(c_p^{\rm m}/T\), where \(c_p^{\rm m} = c_p(\text{Li}_2\text{FeSiO}_4) - 1.01 \times c_p(\text{Li}_2\text{ZnSiO}_4)\). Dashed vertical lines mark \(T_{\rm N}\).}
\label{cp}
\end{figure}

\begin{figure}[htb]
\includegraphics[width=1.0\columnwidth,clip]{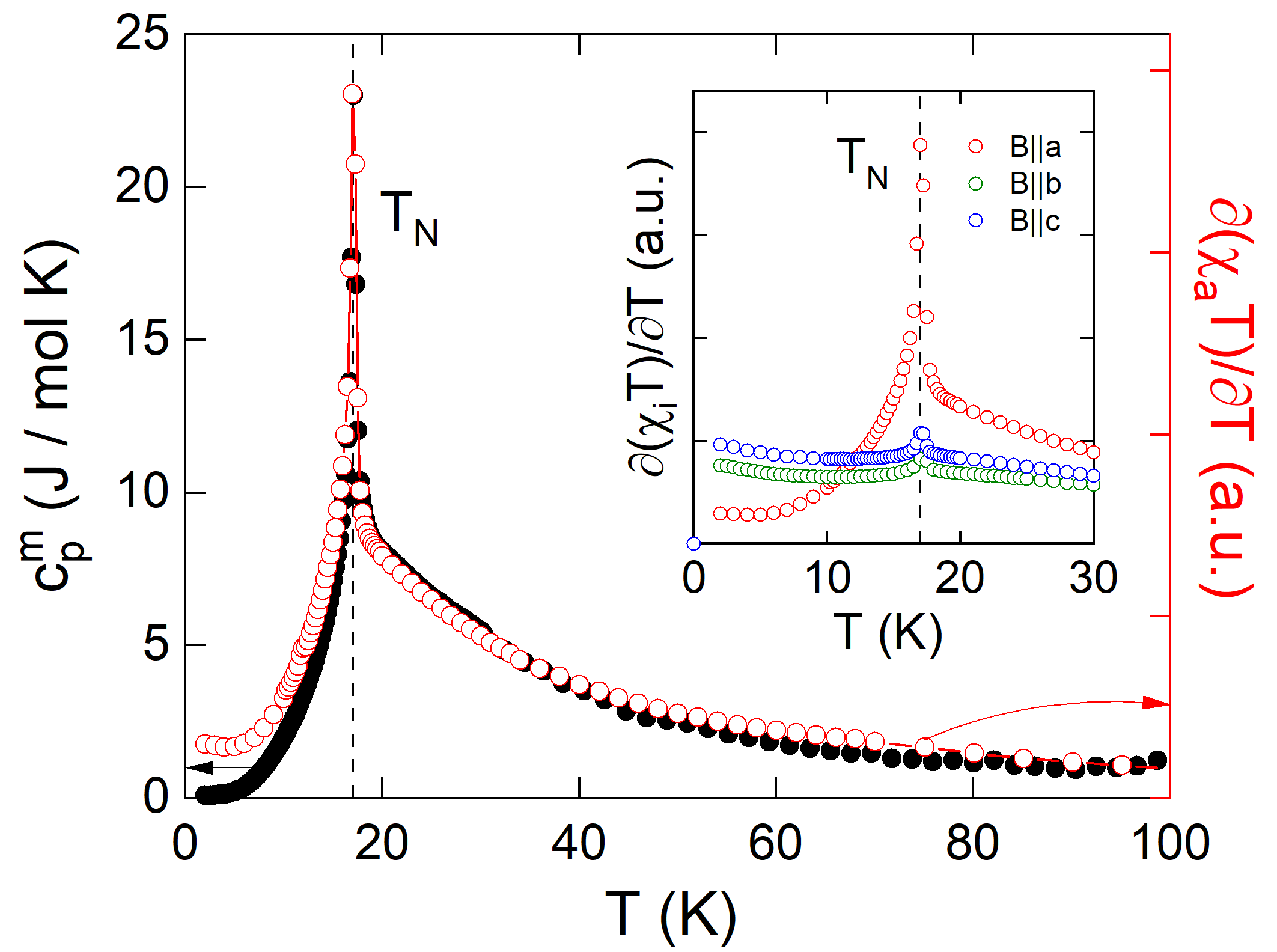}
\caption{Magnetic specific heat, \(c_p^{\rm m}\) (left axis), obtained by subtracting the specific heat of non-magnetic \lzs\ from the data for \lfs\ (see Fig.~\ref{cp}), and the derivative \(\partial (\chi_a T)/\partial T\) of the static magnetic susceptibility, i.e., Fisher's specific heat (right axis). Inset: Fisher's specific heat for all crystallographic axes.}
\label{sm2}
\end{figure}

The comparison of the specific heat of \lfs\ and its compositionally and structurally similar non-magnetic counterpart \lzs, as shown in Fig.~\ref{cp}, demonstrates the predominant phononic nature of entropy changes at high temperatures. The phononic specific heat is superimposed by a pronounced $\lambda$-shaped anomaly which signals a continuous phase transition to a long-range ordered ground state at $T_{\rm N}$. The specific heat of \lzs\ enables assessing the magnetic specific heat \cpm\ = \cpp (\lfs)-1.01*\cpp(\lzs). As shown in Fig.~\ref{sm2}, the so-obtained magnetic specific heat \cpm\ scales excellently with Fisher's specific heat $\partial (\chi_a T)/\partial T$ which confirms the validity of our analysis. Comparison of the data implies significant magnetic entropy changes at least up to 100~K. Integrating \cpm /$T$ yields magnetic entropy changes $\Delta S$($T$), as shown in Fig.~\ref{cp}b. Quantitatively, the obtained magnetic entropy changes saturate at $\Delta S_{\rm tot}\approx 11.7(2)$~\jmk\ which falls within the theoretical prediction $R\ln(2S+1)\approx 13.4$~\jmk . Nearly half of the measured entropy changes appear above \tn, while 6.7~\jmk\ are released in the temperature regime 2~K~$\leq T\leq$~\tn. 

\subsection{Neutron Diffraction}

\begin{figure}[tb]
    \includegraphics[width=1.0\columnwidth,clip]{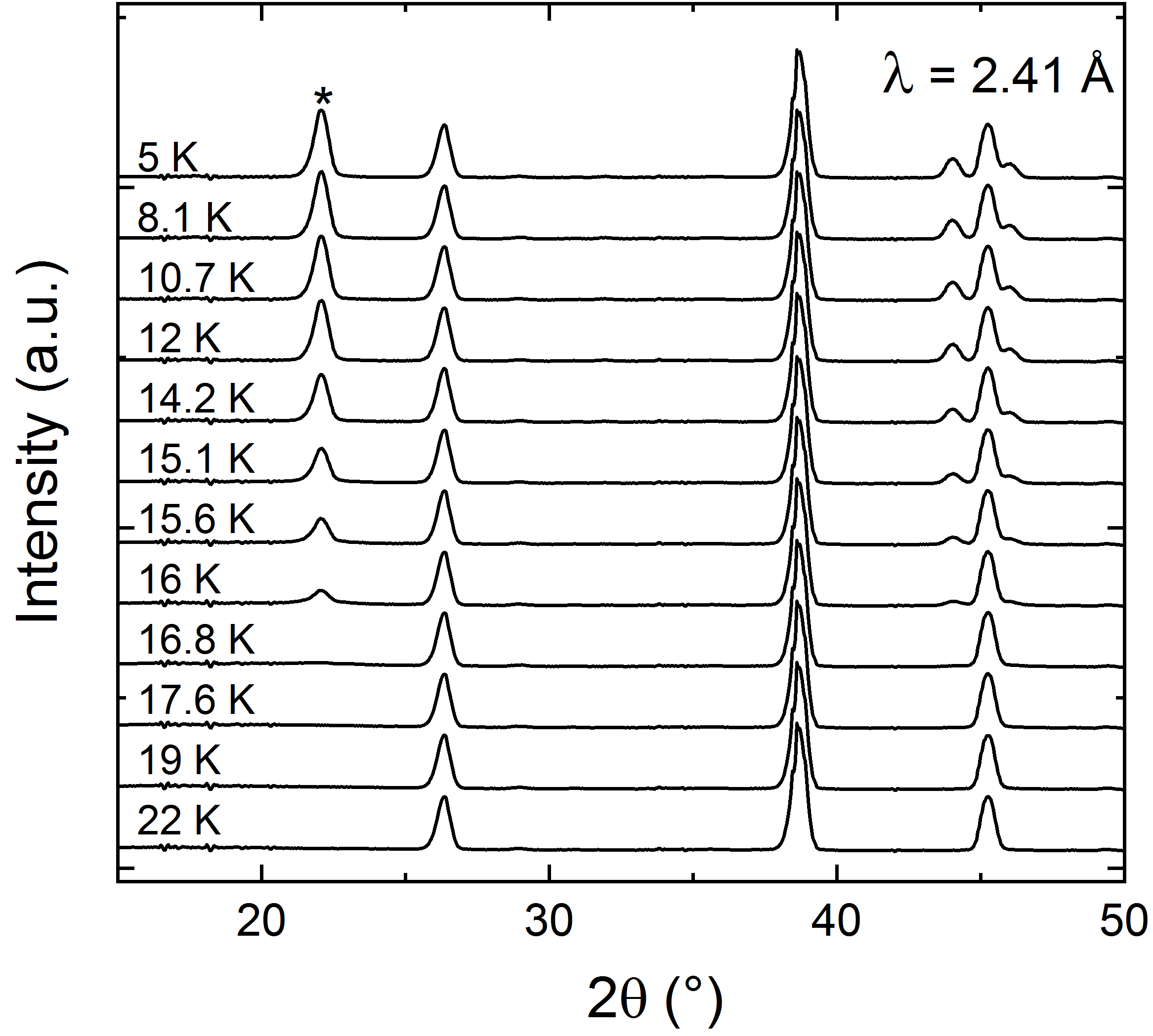}
    \caption{Neutron diffractograms at various temperatures between 5 and 22~K. The asterisk labels the (010) magnetic peak (and equivalents) used for determining the temperature dependence of sublattice magnetization shown in Fig.~\ref{OP}.}
    \label{ND}
\end{figure}

Powder neutron diffraction (PND) in the temperature range between 1.5 and 25\,K was performed to investigate the nuclear and magnetic structure at low temperatures. A detailed description of the full crystal structure determination and refinement of the investigated $Pmnb$-\lfs\ polymorph, achieved via single-crystal X-ray diffraction, is provided in Ref.~\onlinecite{Hergett2019b}. The diffraction profiles in the two-theta range 10\,--\,50$^\circ$ are displayed in Fig.~\ref{ND} for selected temperatures.~\cite{ill} The PND patterns indicate the evolution of superstructure reflections below 16.8\,K. As the temperature decreases, the intensity of the superstructure peaks increases, while their positions remain unchanged in the measured temperature range. The appearance of additional Bragg reflections (one exemplary peak is labeled by the asterisk in Fig.~\ref{ND}) at angles smaller than the angular position of the first nuclear reflection at 26.3$^\circ$ confirms the presence of long-range antiferromagnetic spin order at low temperatures, consistent with the macroscopic data presented above. By indexing the corresponding Bragg reflections, a magnetic propagation vector \textbf{k} = (\nicefrac{1}{2}, 0, \nicefrac{1}{2}) is found. We conclude that the magnetic structure is commensurate with the nuclear lattice. The magnetic unit cell is double the crystallographic one in both the $a$ and $c$-axis directions, while it is the same in the $b$-axis direction. Consequently, there are 16 Fe ions in the magnetic unit cell, while the crystallographic unit cell contains 4 Fe ions. For the space group $Pmnb$ and for \textbf{k} = (\nicefrac{1}{2}, 0, \nicefrac{1}{2}), the magnetic reducible representation $\Gamma_{\rm mag}$ for the Fe$^{\rm 2+}$ (4c) site decomposes as a direct sum of two non-zero irreducible representations (IRs):

\begin{equation}
\label{eq1}
\Gamma_{\rm mag} = 3 \Gamma^{\rm 2}_{\rm 1} \oplus 3 \Gamma^{\rm 2}_{\rm 2}
\end{equation}

\begin{table}[b]
	\centering
	\caption{Basis vectors of the two irreducible representations for \lfs\ with space group $Pmnb$ and \textbf{k} = (\nicefrac{1}{2}, 0, \nicefrac{1}{2}) propagation vector obtained from representational analysis. Fe-1, Fe-2, Fe-3, and Fe-4 refer to the non-primitive basis with coordinates (0.75, 0.42, 0.7), (-0.75, 0.08, 1.2), (1.25, -0.42, -0.7), and (-0.25, 0.92, -0.2), respectively. The magnetic $R$ value ($R_{\rm mag}$) for the refinement based on $\Gamma_{\rm 1}$ yields a satisfactorily low value of 4.73\,\%, while the magnetic ordering schema associated with the $\Gamma_{\rm 2}$ representation is ruled out, as it is incapable of simulating the observed magnetic reflections.}
	
	\begin{tabular}{lcccccc}
		\toprule
		\hline
		IR  & $\Psi_{\rm \nu}$          & Component & Fe-1  & Fe-2  & Fe-3  & Fe-4    \\
		\midrule
		\hline
		\multirow{12}{*}{$\Gamma_{\rm 1}$}& \multirow{2}{*}{$\Psi_{\rm 1}$} & Re  & (100) & (000) & (000) & (-100)  \\
		&                                 & Im  & (000) & (100) & (-100)& (000)   \\
		& \multirow{2}{*}{$\Psi_{\rm 2}$} & Re  & (010) & (000) & (000) & (010)   \\
		&                                 & Im  & (000) & (010) & (010) & (000)   \\
		& \multirow{2}{*}{$\Psi_{\rm 3}$} & Re  & (001) & (000) & (000) & (00-1)  \\
		&                                 & Im  & (000) & (00-1)& (001) & (000)   \\
		& \multirow{2}{*}{$\Psi_{\rm 4}$} & Re  & (000) & (-100)& (100) & (000)   \\
		&                                 & Im  & (-100)& (000) & (000) & (100)   \\
		& \multirow{2}{*}{$\Psi_{\rm 5}$} & Re  & (000) & (010) & (010) & (000)   \\
		&                                 & Im  & (010) & (000) & (000) & (010)   \\
		& \multirow{2}{*}{$\Psi_{\rm 6}$} & Re  & (000) & (00-1)& (001) & (000)   \\
		&                                 & Im  & (001) & (000) & (000) & (00-1)  \\
		
		\hline
		\addlinespace
		\multirow{12}{*}{$\Gamma_{\rm 2}$}& \multirow{2}{*}{$\Psi_{\rm 1}$} & Re  & (100) & (000) & (000) & (100)   \\
		&                                 & Im  & (000) & (100) & (100) & (000)   \\
		& \multirow{2}{*}{$\Psi_{\rm 2}$} & Re  & (010) & (000) & (000) & (0-10)  \\
		&                                 & Im  & (000) & (010) & (0-10)& (000)   \\
		& \multirow{2}{*}{$\Psi_{\rm 3}$} & Re  & (001) & (000) & (000) & (001)   \\
		&                                 & Im  & (000) & (00-1)& (00-1)& (000)   \\
		& \multirow{2}{*}{$\Psi_{\rm 4}$} & Re  & (000) & (100) & (100) & (000)   \\
		&                                 & Im  & (100) & (000) & (000) & (100)   \\
		& \multirow{2}{*}{$\Psi_{\rm 5}$} & Re  & (000) & (0-10)& (010) & (000)   \\
		&                                 & Im  & (0-10)& (000) & (000) & (010)   \\
		& \multirow{2}{*}{$\Psi_{\rm 6}$} & Re  & (000) & (001) & (001) & (000)   \\
		&                                 & Im  & (00-1)& (000) & (000) & (00-1)  \\
		\bottomrule
		\hline
		
	\end{tabular}
	\label{irs}
\end{table}

\begin{figure}[bt]
	\includegraphics[width=0.8\columnwidth,clip] {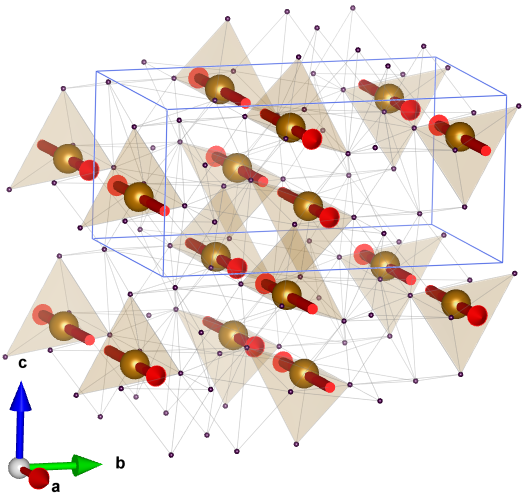}
	\caption{Sketch of the spin configuration. Lines show the crystallographic unit cell. }\label{moments}
\end{figure}

The basis vectors of these IRs are listed in Tab.~\ref{irs}. Of the two allowed antiferromagnetic spin configurations, only $\Gamma_{\rm 1}$ can reproduce the measured magnetic intensities. The resulting spin configuration is visualized in Fig.~\ref{moments}. The magnetic moments of Fe$^{\rm 2+}$ are aligned antiferromagnetically along the $a$-axis with an ordered moment of 2.50(2)\,\mb /Fe at $T = 1.5$\,K. Its magnitude is smaller than the expected value of $\sim 4.5$\,\mb\ for magnetic moment of Fe$^{\rm 2+}$ (considering the measured $g=2.27$) in the high-spin $S=2$-state.

\begin{figure}[tb]
\includegraphics[width=1.0\columnwidth,clip]{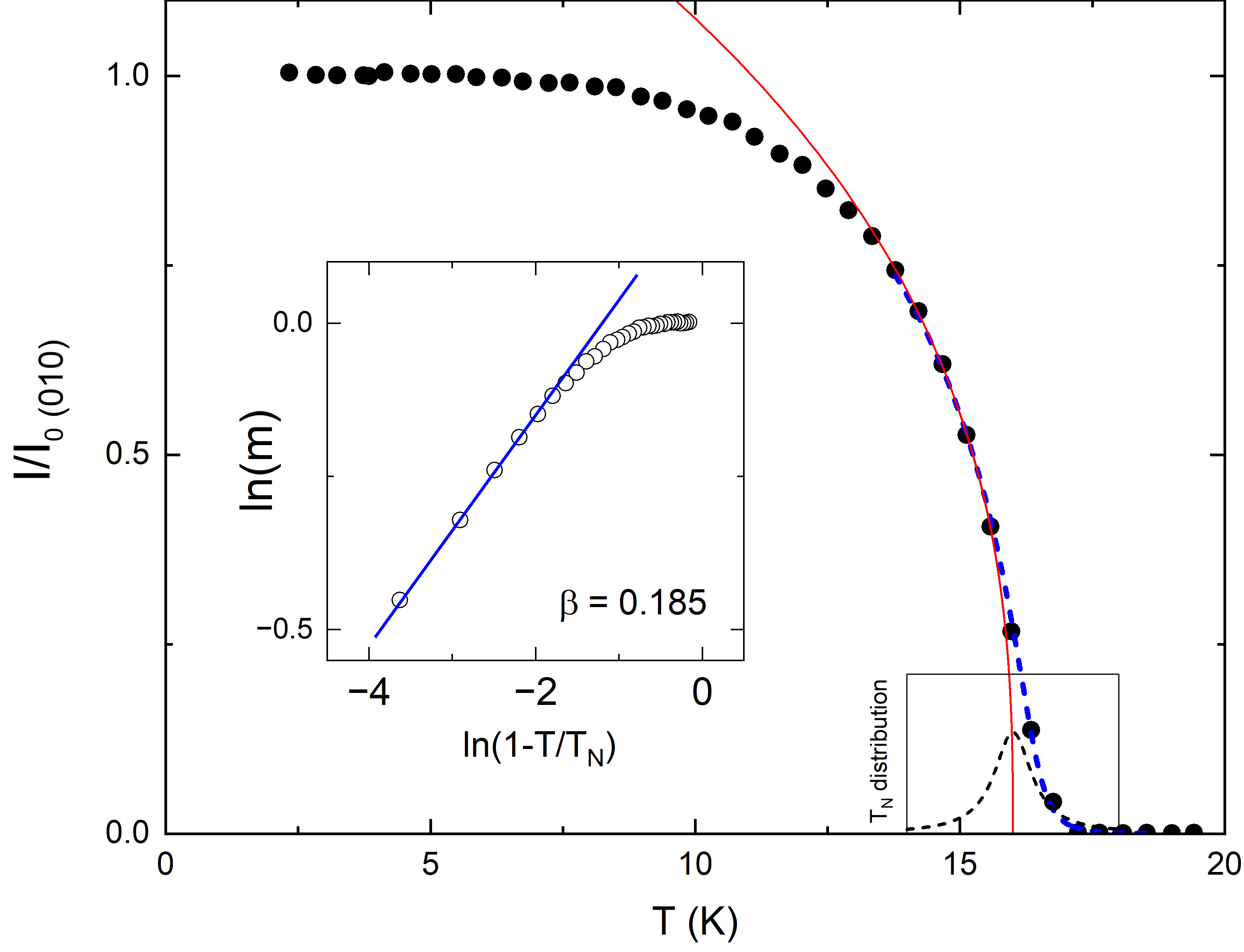}
\caption{Order parameter fit analysis of the intensity of the (010) magnetic superstructure reflection (see Fig.~\ref{ND}). The solid red line represents a fit to the data by a power-law and the blue dashed line the fit by Eq.~\ref{eq2}, i.e., by considering the variation of the N\'{e}el temperature as shown by the black short-dashed line
(see the text)}. The inset shows a logarithmic plot of the sublattice magnetization as a function of the reduced temperature. The solid blue line in the inset represents a fit obtained with the critical exponent $\beta = 0.185$.
	\label{OP}
\end{figure}

Fig.~\ref{OP} shows the integrated intensity ($I_{B} \propto |M_{s}|^2$, where $M_{s}$ is the order parameter) of the strongest magnetic peak (010) in the temperature range 2~--~19\,K. The intensity vanishes around $T_{\rm N}$, which agrees with the specific heat data. In the conventional picture of a continuous phase transition, the magnetic order parameter follows a power-law equation. By fitting the integrated intensity to the power-law scaling function $I_{B} = I_{0} |t|^{2\beta}$, where $t = 1 - T/T_{\rm N}$ is the reduced temperature, an estimate for the critical exponent $\beta = 0.18(2)$ is obtained.\footnote{For fitting the data, we used $T_{\rm N} = 16$\,K, as determined from characterizing the powder sample used for the neutron study.} As will be discussed below, we interpret the thus obtained power-law exponent only as an effective value describing a convolution of temperature dependence of the order parameter and of the magnetic volume fraction, i.e., it does not directly reflect the critical behaviour (see §~\ref{sec:MB}). 
 
In the vicinity of \tn\ the temperature dependence of the order parameter is not very well described by a power-law which would yield a sharp kink at \tn\ instead of the observed smeared-out behaviour (see Fig.~\ref{OP}). A better fit is obtained by assuming a Lorentzian distribution of \tn\ which may result from strain effects or tiny variations of oxygen content in the sample. Assuming the distribution of \tn\ with full width at half maximum (FWHM) around a center N\'{e}el temperature $T^{\rm cen}_{\rm N}$ in the measured polycrystal, the fit to the data is obtained by performing a convolution of the power-law scaling function

\begin{equation}
 \label{eq2}
f(T_{\rm N}) = f_{0} \bigg[ \frac{\gamma}{(T_{\rm N} - T^{\rm cen}_{\rm N})^2 + \gamma^2} \bigg].
\end{equation}
Here, $f_{0}$ is a normalization prefactor, and $\gamma$ is the scale parameter, which determines the FWHM = 2$\gamma$.
The resultant fit is shown by the dashed blue line in Fig.~\ref{OP}, with $T^{\rm cen}_{\rm N} = (16.02 \pm 0.02)$\,K and $\gamma = (0.43 \pm 0.05)$\,K.

\subsection{M\"ossbauer spectroscopy}\label{sec:MB}

\begin{figure}[!]
\includegraphics[width=0.5\columnwidth,clip]{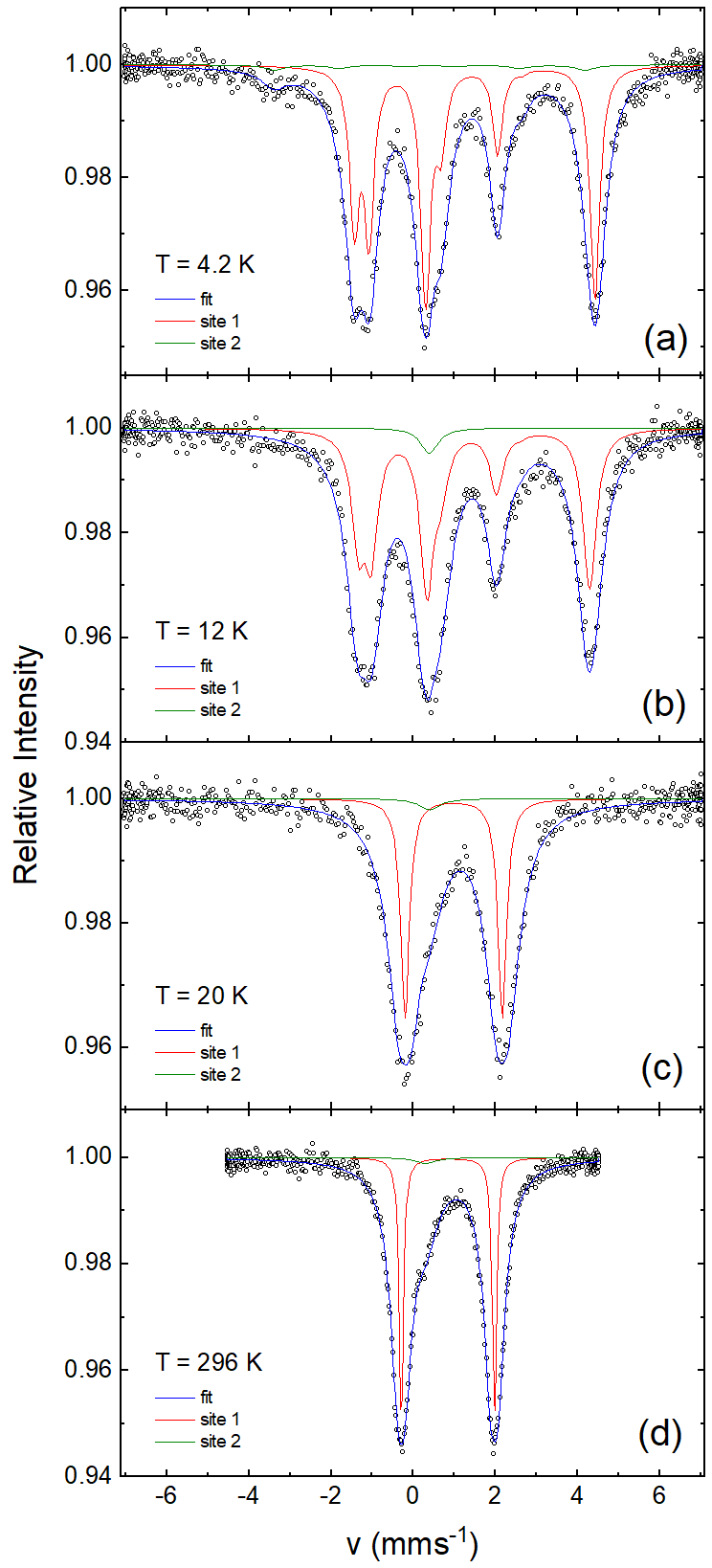}
\caption{Typical Mössbauer spectra of \lfs\ at 4.2, 12, 20 and 296~K. The spectra can be described by a static hyperfine Hamiltonian analysis (blue) assuming two magnetic sites. The main site (red) exhibits an electric field gradient as well as a static hyperfine field below 20~K. Site 2 contributes with $\approx 5$~\%. It exhibits a singlet above 4.2~K and shows a static magnetic hyperfine field at 4.2~K.}
\label{fig:Moesspec}
\end{figure}

Representative Mössbauer spectra of \lfs\ are shown in  Fig.~\ref{fig:Moesspec}. All spectra are analysed using a static nuclear hyperfine Hamiltonian for powder samples.
At $T=4.2$~K, two Fe sites are observed. The main site has an intensity fraction of $a_1 = 95.6(6)$~\% and is characterized by an electric field gradient (EFG) as well as a magnetic hyperfine field of \Bhyp~=~14.8(3)~T. The principal EFG component of $V_{zz} = -127.3(8)$~V/\angstrom$^2$ is orientated orthogonal to the magnetic hyperfine field, which is parallel to its smallest principal axis $V_{yy}$ with the three principle axis of the EFG given by $|V_{zz}| \geq |V_{yy}| \geq |V_{xx}|$. The EFG shows an asymmetry parameter $\eta = (V_{yy} - V_{xx})/V_{zz} = 0.752(8)$. The main site exhibits an isomer shift $\delta = 1.109(10)$~mm/s. The second site with an relative intensity of $a_2 = 4.4(6)$~\% shows a magnetic hyperfine field of \Bhyp~=~23.5(10)~T with an isomer shift of 0.5(2)~mm/s.

At room temperature, the spectra imply the presence of two sites, too (see Fig.~\ref{fig:Moesspec}d). For the main site (relative intensity $a = 94.7(30)$~\%) the asymmetry parameter $\eta$ obtained from the low temperature spectrum at 4.2\,K is assumed. This site shows a quadruple-splitting with an EFG principal component of $V_{zz} = -124.8(20)$~V/$\angstrom^2$, similar to the data at 4.2~K. The isomer-shift is $\delta= 0.96(1)$~mm/s. The minority site with a relative intensity of $a = 5.3(30)$~\% is described by a broadened singlet (\textit{i.e.}, EFG~$\approx 0$) with an isomer shift of $\delta = 0.42 (10)$~mm/s. This site may be associated with excess iron ions on an interstitial position in the unit cell. 
	
The isomer shift of $\delta= 0.96(1)$~mm/s for the main site at room temperature in combination with the EFG value unambiguously confirm the Fe$^{2+}$ $(S=2)$ high-spin state in \lfs . The finite value of the EFG asymmetry parameter $\eta$ is in agreement with the distorted tetrahedral crystal field around the Fe nucleus.~\cite{Hergett2019b}
		
Between room temperature and 20~K the M\"ossbauer spectra show no static hyperfine field which is consistent with a paramagnetic state. Below 20~K a magnetic phase transition is observed via the gradual appearance of a static magnetic hyperfine field $B_{\rm hyp}$. For all temperatures below \tn\ the angle $\vartheta$ between $V_{\rm zz}$ and $B_{\rm hyp}$ determined from the M\"ossbauer spectra analysis is 90(2) degrees. The orientation of $V_{zz}$ orthogonal to $B_{\rm hyp}$ is in agreement with the iron magnetic moment orientation parallel to the undistorted tetrahedral axis as deduced from macroscopic magnetisation and neutron powder diffraction as discussed above.

\begin{figure}[tb]
\includegraphics[width=1.0\columnwidth,clip]{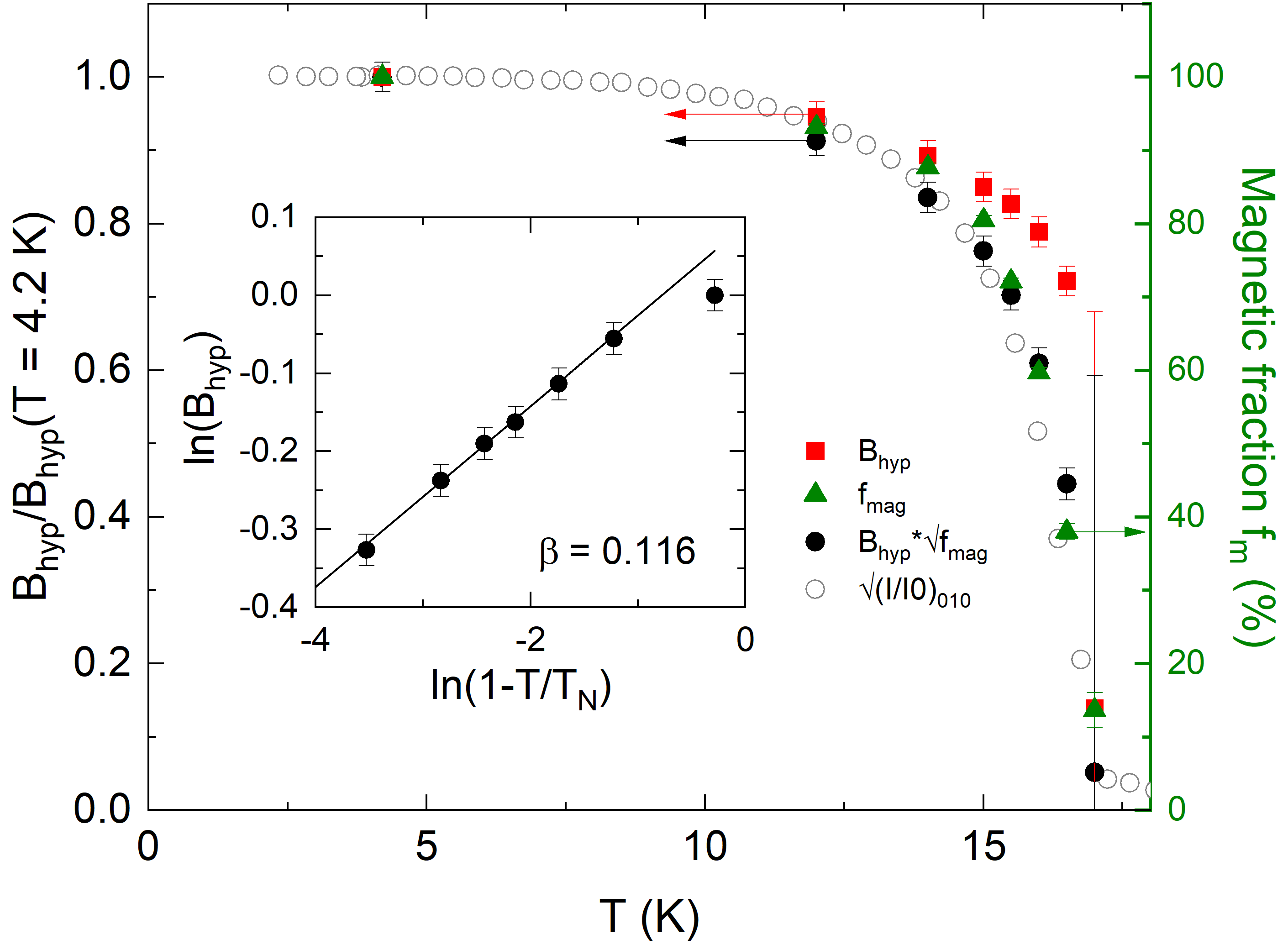}
\caption{Temperature dependence of the normalised magnetic hyperfine field $B_{\rm hyp}$ below 20~K (red squares, left ordinate) and of the magnetic volume fraction $f_{\rm mag}$ determined from the analysis of the Mössbauer spectra (green triangles, right ordinate). The black solid circles describe the product $B_{\rm hyp}\times \sqrt{f_{\rm mag}}$, the black open circles describe $\sqrt{I_{010}}$) from Fig.~\ref{OP} (all normalised to their low-temperature values). The inset shows a double-logarithmic plot of $B_{\rm hyp}$ as a function of the of the reduced temperature.}
\label{fig:Bhyp}
\end{figure}

The temperature dependence of $B_{\rm hyp}$, which is proportional to the magnetic order parameter $M_{s}$, is shown in Fig.~\ref{fig:Bhyp} (solid red squares). For comparison, we include the scaled temperature dependence of the neutron magnetic order parameter $M_{s}$ from Fig.~\ref{OP} (black open circles). Below 13\,K both data sets are following the same trend. However, between 14 and 17\,K the M\"ossbauer hyperfine field $B_{\rm hyp}$ is always above the corresponding neutron value.

This discrepancy is not expected for three-dimensional magnets (see, e.g., Ref.~\onlinecite{Koebler2015} and references therein). In the local probe M\"ossbauer spectroscopy the determination of the temperature dependent magnetic order parameter $B_{\rm hyp} (T)$ is independent of the magnetic volume fraction $f_{\rm mag}(T)$ determined from the signal intensity. In contrast, the neutron diffraction magnetic order parameter obtained from the intensity of a magnetic Bragg peak, i.e., $\sqrt{I_{\rm 010}(T)}$ is proportional to the static magnetic dipole moment $M_{\rm s}(T)$ multiplied by the square root of $f_{\rm mag} (T)$. The continuous decrease of $f_{\rm mag} (T)$ from 1 to 0 in the vicinity of the magnetic phase transition (determined from the analyis of the M\"ossbauer spectra and shown in Fig.~\ref{fig:Bhyp}, right ordinate) leads to the suppression of the neutron magnetic order parameter in this temperature range. For illustration in Fig.~\ref{fig:Bhyp} we also include a plot of the product $B_{\rm hyp}(T) \times \sqrt{f_{\rm mag}(T)}$ (filled black circles). These data are in very good approximation proportional to the experimental values of $\sqrt{I_{\rm 010}(T)}$.  

The inset of Fig.~\ref{fig:Bhyp} shows a logarithmic plot of $B_{\rm hyp} (T)$ as a function of the reduced temperature. The solid blue line represents a fit obtained with the critical exponent $\beta = 0.116(12)$ and $T_{\rm N}^{\rm MB}=17$~K. Since $B_{\rm hyp} (T)$ is directly proportional to $M_{\rm s}(T)$ this value of $\beta$ may be considered as the thermodynamic critical exponent of the magnetic phase transition in \lfs .

Next we discuss the absolute value of the iron Mössbauer magnetic hyperfine field $B_{\rm hyp} = 14.8(3)$~T at the lowest measured temperature of $T=4.2$~K. This value is consistent with the absolute value of the ordered dipole moment of 2.5~$\mu_{\rm B}$ as determined by our neutron diffraction experiment. In a magnetic insulator $B_{\rm hyp}$ is given by~\cite{hyperfineinteraction}
 
\begin{equation}
B_{\rm{hyp}} = B_{\rm s} + B_{\rm L} + B_{\rm D},
 	\end{equation}
  
where $B_{\rm s}$ is the Fermi contact field being typically in the range of 20 -- 50~T for high spin Fe$^{2+}$. $B_{\rm L}$ is the orbital field which is usually of the order of 20~T and anti-parallel to $B_{\rm s}$.
$B_{\rm D}$ is the dipole field usually at the order of a few tesla for iron.~\cite{hyperfineinteraction} The measured $B_{\rm{hyp}}$ hence suggests the presence of a measurable orbital contribution to the local hyperfine field and hence to the magnetic order parameter.

\begin{table}[h]
\caption{Comparison of M\"ossbauer hyperfine parameters for \lfs\ (this work) with those of Cu$_2$FeGeS$_4$ (Ref.~\onlinecite{imbert}). \Bhyp , $V_{zz}$, $\vartheta$, and $\eta$ are the magnetic hyperfine field, the principal component of the EFG, the angle between $V_{\rm zz}$ and $B_{\rm hyp}$, and the asymmetry parameter as discussed in the text.}
\label{tab:hypint}
\begin{tabular}[c]{c|c|c|c|c|c|c}
		 & \Bhyp [T] & $V_{zz}$ [V/\AA$^2$] & $\vartheta$ [$\degree$] & $\eta$\\ \hline
		Li$_2$FeSiO$_4$ &14.8(3) & -127.3(8)& 90(2) & 0.752(8) \\
		Cu$_2$FeGeS$_4$ & 16.7(2) & -151.8(18) & 90(1) & 0.00(3) \\
	\end{tabular}
        \end{table}

In Tab.~\ref{tab:hypint} the Mössbauer hyperfine parameters of \lfs\ are compared to those for high spin Fe$^{2+}$ in the structurally similar compound Cu$_2$FeGeS$_4$ analyzed in Ref.~\onlinecite{imbert}. In both systems, magnetism is associated with tetrahedrally coordinated iron sites. A main difference between both materials is that the tetrahedra in Cu$_2$FeGeS$_4$ are not distorted in contrast to what is observed in \lfs.~\cite{Hergett2019b} Accordingly, $\eta$ is found to vanish in Cu$_2$FeGeS$_4$ whereas the distorted tetrahedral environment in the $bc$ plane of \lfs\ results in $\eta= 0.752(8)$. 	

\section{Numerical studies}

\begin{figure}[b]

\centering
\includegraphics[width=0.5\columnwidth,clip]{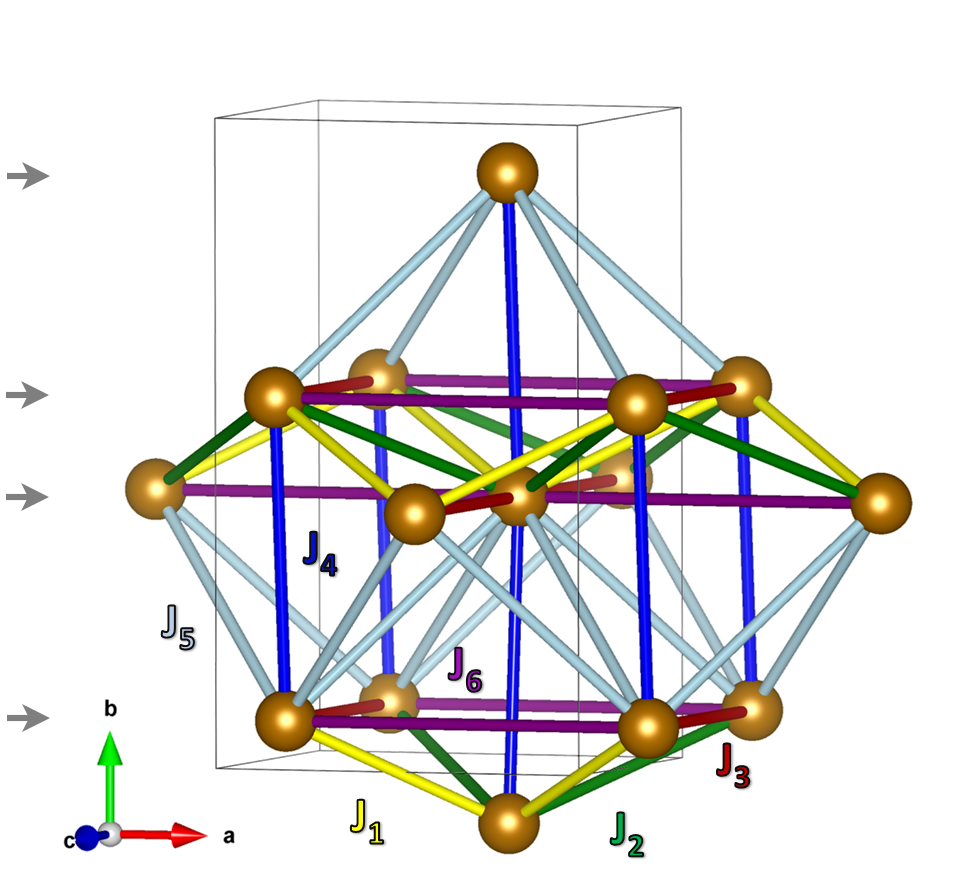}
\caption{Illustration of the magnetic interactions between Fe$^{2+}$ moments which have been considered in the model (up to the 6th nearest neighbor). Thin black lines show the crystallographic unit cell. Four horizontal grey arrows on the left indicate the successive Fe-layers stacked along the $b$ direction within the shown unit cell.}
\label{couplingconstants}
\end{figure}

The magnetic coupling parameters were obtained by comparison of the DFT+U calculated total energies for several magnetic configurations. This approach, sometimes called 'broken-symmetry formalism'~\cite{Saul_magnetic_2011,Saul2014}, is quite common in the literature.~\cite{Zvereva_zigzag_2015,Foyevtsova_determination_2011} Individual exchange couplings $J_{l}$ were defined by the spin Hamiltonian $\sum_{\langle i ,j \rangle  } J_l \mathbf{S_i} \mathbf{S_j}$, where $i$ and $j$ are $l$th~neighbors, $\mathbf{S_i}$ are the spin operators located on site $i$ divided by $\hbar$ ($\mathbf{S_i}$ is dimensionless) and $J_l$ are the magnetic exchange coupling between $l$th neighbors ($J_l$ are energies). Here, we have included six nearest neighbors as depicted in Fig.~\ref{couplingconstants}. The corresponding distances are listed in Table~\ref{tab:J}.

We studied the ferromagnetic (FM) configuration and 21 different antiferromagnetic (AFM) configurations that we label $\alpha \in [1,\hdots,21]$. All DFT+U self-consistent field calculations converged to structures in which the magnetic moment per atom is $1.87 \mu_{B}$ at all sites. For a given configuration the Heisenberg Hamiltonian can therefore be rewritten as a linear combination of $J_l$ with dimensionless coefficients $c_l^\alpha$ which only depend on the spins on the bond being parallel ($c_l=1$) or anti-parallel ($c_l=-1$) in the configuration:
\begin{equation}
-\sum_{\mathclap{\substack{\langle i ,j \rangle \\ l-th~neighbor}}} J_l \mathbf{S_i} \mathbf{S_j}=-\sum_{l=1}^{6}  J_l^{\rm{DFT}} c_l^\alpha .
\label{Eq:HeisenbergModel}
\end{equation}
For the ferromagnetic configuration, all the $c_l$ are equal to +1 while for AF structures they are either +1 or -1 depending on the relative spin orientation on the bond.

\begin{figure}[htb]
	\includegraphics[width=1.0\columnwidth]{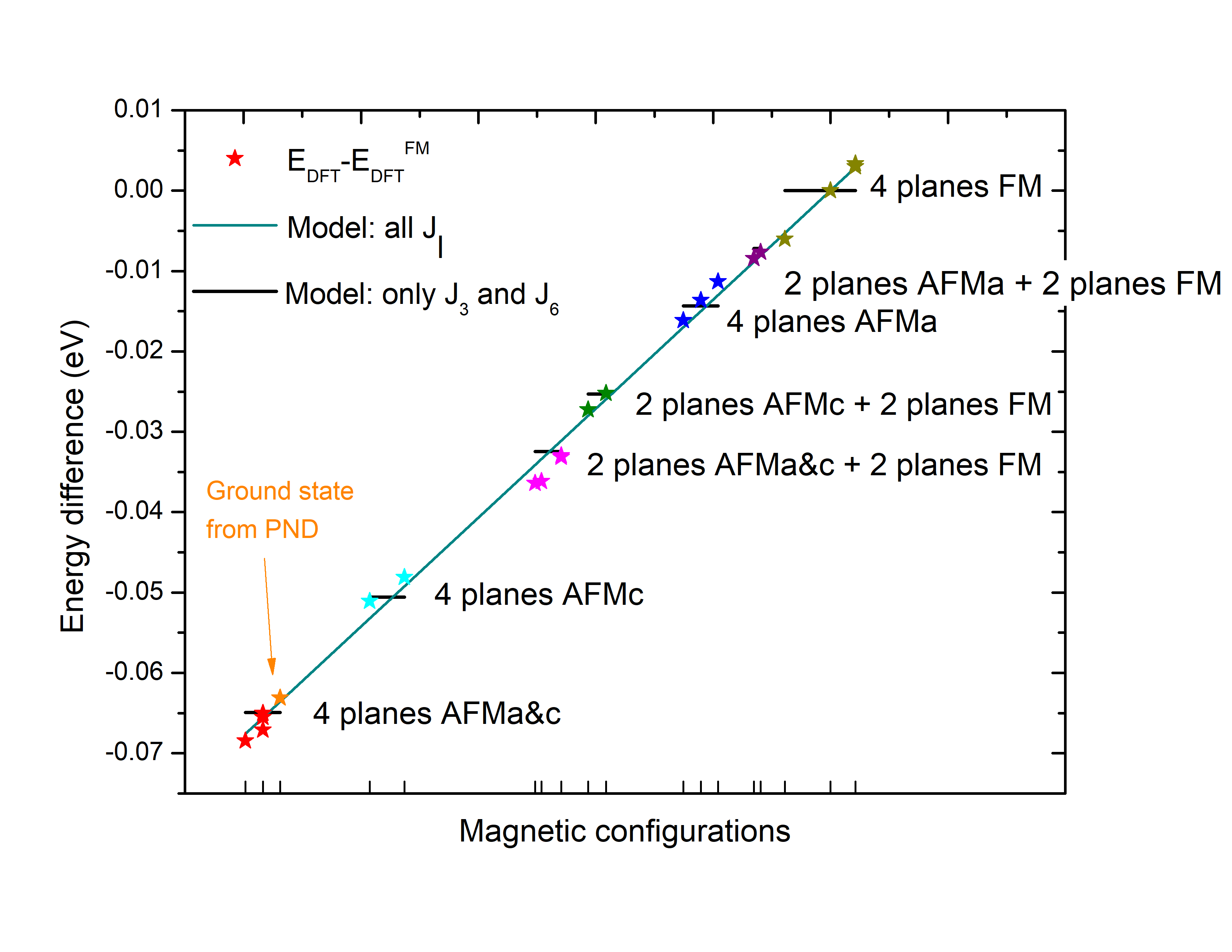}
	\caption{Total energies computed for several antiferromagnetic spin configurations compared with the values obtained with the Heisenberg model Eq.~\eqref{Eq:HeisenbergModel}. For clarity, magnetic configurations are spread along the $x$ axis so that $\sum_{l=1}^{6} c_l^\alpha  J_l^{\rm{\rm DFT}} $ appears to be linear. The star scatters depict the DFT calculated total energy for a given magnetic structure minus the calculated total energy for the ferromagnetic structure. The cyan line corresponds to $\sum_{l=1}^{6} -(c_l^\alpha -1)  J_l^{\rm DFT}$ with the values for $J_l$ listed in table~\ref{tab:J}. The black lines depict the model Hamiltonian considering only $J_3$ and $J_6$ as non-zero. Abbreviations (see also Fig.~\ref{plans}): AFMi(\&j) = AF alignment of neighboring moments in direction(s) $i$ (\&$j$) and FM alignment along the other in-plane direction.}
 
	\label{Energy_differences}
\end{figure}

In Fig.~\ref{Energy_differences}, the energy $E_{\rm DFT}^\alpha -E_{\rm DFT}^{\rm FM}$ is plotted against $\sum_{l=1}^{6} (c_l^\alpha -1) J_l M_S^2$ for the 21 different antiferromagnetic configurations. The calculated energy of the experimental antiferromagnetic configuration as determined by neutron diffraction is represented in orange. In our calculations, it is almost the lowest energy configuration. 
Using the LeastSquares function of Mathematica,\cite{Mathematica} we determined the values of $J_l$ that minimize the quantity:
\begin{equation}
\sum_\alpha \left[ (E_{\rm DFT}^\alpha - E_{\rm DFT}^{\rm FM} ) - \sum_{l=1}^{6} (c_l^\alpha -1) J^{\rm{\rm DFT}} \right]^2
\end{equation}
where $E_{\rm DFT}^\alpha$ is the calculated total energy for the configuration $\alpha$ and $E_{\rm DFT}^{\rm FM}$ is the calculated total energy for the ferromagnetic configuration.
Fig.~\ref{Energy_differences} also shows (as a line) the function  $E = \sum_{l=1}^{6} (c_l^\alpha -1) J_l M_S^2 $ which illustrates the quality of the fitted exchange parameters $J_l$.

The obtained isotropic exchange couplings $J_l$ are listed in Table~\ref{tab:J}. The largest values are in the meV range, which is consistent with the experimentally observed magnetic transition temperature \tn . 

\begin{table}[htb]
	\caption{Coupling constants and corresponding Fe-Fe distances $d_{\rm Fe-Fe}$: $J^{\rm{DFT}}$ are the raw solutions in meV and $J_L=J^{\rm{DFT}} /S(S+1)=J^{\rm{DFT}} /6 $ are the actual coupling constants. It is important to note that, because of the driving role of $J_3$ and $J_6$ (see text), the values obtained for all other $J_l$ are subject to a large relative uncertainty.}
	\label{tab:J}
	\begin{tabular}{@{\extracolsep{4pt}}llllc}
		\hline
		\hline
		~ & $d_{\rm{Fe-Fe}}$  & & $ J^{\rm{\rm DFT}}$(meV)  & $J_L $ (K)\\
		\hline
		$J_1$ & 4.114 $\angstrom$ &                        &  -2.2$\times 10^{-3}$ & -0.004 K\\
		$J_2$ & 4.684 $\angstrom$ &                        &  -2.6$\times 10^{-1}$  &  -0.50 K \\
		$J_3$ & 5.031 $\angstrom$ & $||~\mathbf{c}$        &  -3.2  &  -6.1 K  \\
		$J_4$ & 5.338 $\angstrom$ & almost $||~\mathbf{b}$ &  2.5$\times 10^{-1}$ & 0.48 K \\
		$J_5$ & 5.387 $\angstrom$ &                        &  -3.5$\times 10^{-2}$  &  -0.068 K   \\
		$J_6$ & 6.278 $\angstrom$ &  $||~\mathbf{a}$       &  -0.9&  -1.74 K \\
		\hline
		\hline
	\end{tabular}
\end{table}

Except $J_4$, all considered interactions are antiferromagnetic (negative $J_l$). The magnetic interactions in the  $\mathbf{a}$ and $\mathbf{c}$ directions are the largest one ($J_3$ along $\mathbf{c}$ and $J_6$ along $\mathbf{a}$) which suggests that the magnetic configurations inside the $(\mathbf{a},\mathbf{c})$ plane drive the magnetic properties of the crystal. 
This is supported by Fig.~\ref{Energy_differences}. Indeed, the calculated energies (colourful scatters) are gathered into groups that correspond to different $(\mathbf{a},\mathbf{c})$-plane configurations, independently of the ordering in the $\mathbf{b}$ direction. 

\begin{figure}[htb]
\includegraphics[scale=0.5]{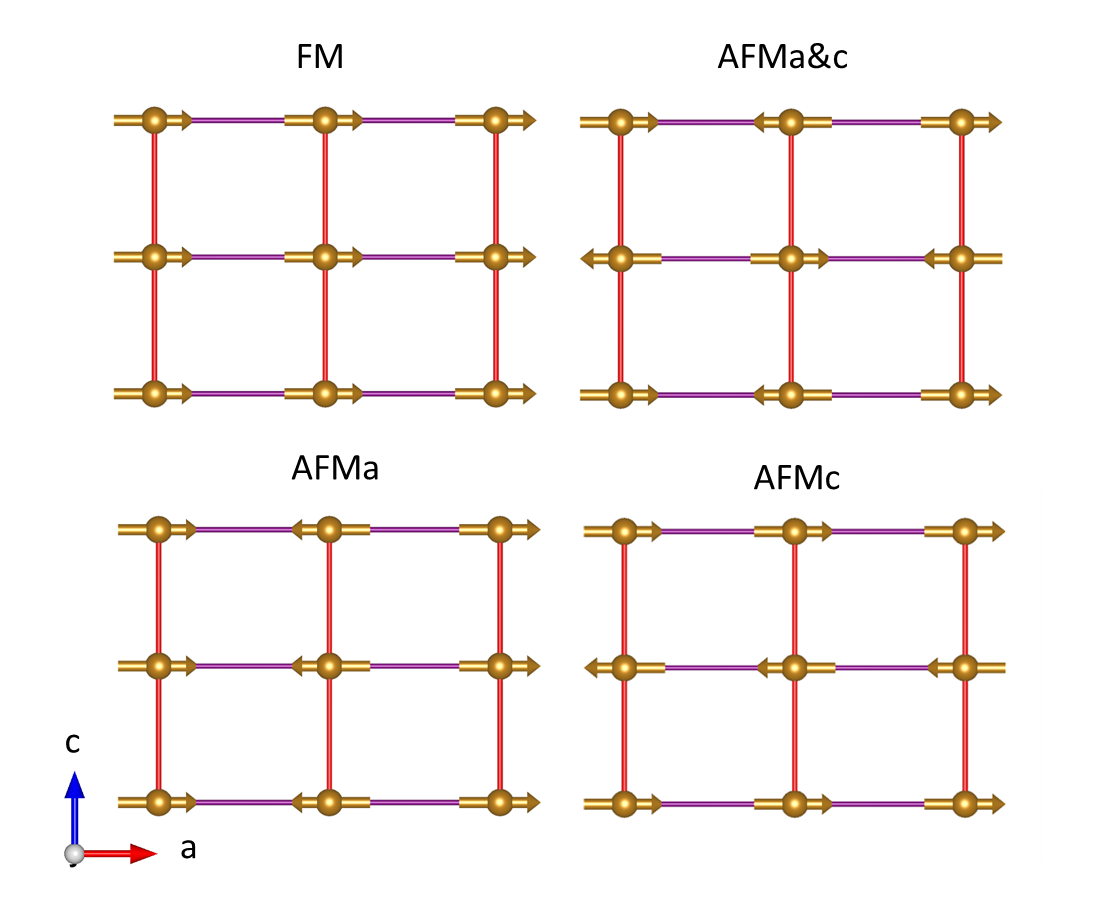}
\caption{Several possible magnetic configurations in the $(\mathbf{a},\mathbf{c})$ plane. The red and purple lines correspond respectively to the interactions $J_3$ and $J_6$. }
\label{plans}
\end{figure}

The different magnetic patterns in the  $(\mathbf{a},\mathbf{c})$-planes are depicted in Fig.~\ref{plans}. As $J_3$ and $J_6$ are both antiferromagnetic and $|J_3| > |J_6|$, we expect
$
E_{\rm AFMc\&a}< E_{\rm AFMc}< E_{\rm AFMa} < E_{\rm FM}.
$
The energy order observed in Fig.~\ref{Energy_differences} is perfectly consistent with this in-plane energy order. 

In addition, we have evaluated an alternate Heisenberg model with all $J_l$ except $J_3$ and $J_6$ (i.e., the $ac$-planar exchange couplings) set to zero (see the black lines in Fig.~\ref{Energy_differences}). Notably, this purely planar model is sufficient to describe the different families of magnetic configurations with an acceptable accuracy. In other words, \lfs\ exhibits almost two-dimensional magnetism.

\section{Discussion and Summary}

While the evolution of long-range antiferromagnetic order in \lfs\ at $T_{\rm N} = 17$~K is marked by sharp $\lambda$-like anomalies in the specific heat and $\partial (\chi_a T) / \partial T$, our experimental data also suggest the presence of short-range spin order at temperatures above $10\times T_{\rm N}$. The presence of a broad maximum in the magnetic susceptibility well above the long-range ordering temperature is typical for low-dimensional antiferromagnetism and indicates the evolution of significant short-range magnetic order~\cite{dejongh1974}. Both the observation of such a correlation maximum at 28~K and the fact that the Weiss temperature $\Theta$ exceeds $T_{\rm N}$ by a factor of $\simeq 3.5$ ($T_{\rm N}/\Theta \simeq 0.27(2)$) suggest the low-dimensional nature of magnetism in \lfs. This is supported by the observation of a reduced ordered moment $\mu = 2.50(2)$~\mb /Fe, at $T = 1.5$~K, obtained from neutron diffraction. Additionally, we find that magnetic anisotropy beyond $g$-factor anisotropy extends to temperatures up to about 200~K, i.e., more than $10\times T_{\rm N}$, as well as significant non-phononic entropy changes up to at least 100~K. Since strong single-ion effects are rather unlikely in the high-spin Fe$^{2+}$ system under study, we associate the observed anisotropy in $\chi$ with the evolution of short-range correlations presumably of magnetic nature. However our data do not exclude effects of orbital degrees of freedom either. The presence of short-range order up to at least 100~K is unambiguously evidenced by the measured non-phononic entropy changes. At 100~K, the magnetic entropy is still not fully released which agrees with the scenario of short-range magnetic order up to 200~K. The presence of short-range magnetic order well above $T_{\rm N}$ further corroborates the evidence of low-dimensional magnetism in \lfs.

This conclusion of low-dimensional magnetism is fully confirmed by our numerical studies, where the magnetic couplings were estimated using the broken-symmetry formalism by minimizing total energies of various spin configurations and mapping onto the Heisenberg hamiltonian. The experimentally observed antiferromagnetic spin configuration, as determined by our neutron diffraction experiment, is indeed found to exhibit nearly the lowest energy, supporting the validity of our approach. Our analysis yields as dominating magnetic exchange couplings $J_3=-6.1$~K ($||c$ axis) and $J_6=-1.7$~K ($||a$ axis) forming two-dimensional magnetic layers.  While $J_1$ and $J_5$ are negligibly small, $J_2$ and $J_4$ provide finite interplanar coupling which, in combination with magnetic anisotropy, present derivations from the pure 2D Heisenberg case and, hence, drive long-range magnetic order. The antiferromagnetic interplane coupling $J_2=-0.5$~K between adjacent layers of Fe$^{2+}$ moments is frustrated so that the next-nearest neighbor coupling $J_4=0.48$~K is given as leading interplane coupling. It is ferromagnetic which contradicts the observed spin structure. However, the error in the calculated value is rather large (due to the crushing driving role of $J_3$ and $J_6$ in the coupling) so that even its sign is subject to uncertainty. Hence, our numerical study does not rule out a small negative $J_4$ which would be consistent with the observed superstructure vector \textbf{k}=(\nicefrac{1}{2},0,\nicefrac{1}{2}).

The high-spin nature of Fe$^{2+}$ in \lfs\ is further confirmed by our Mössbauer data which imply a main magnetic Fe$^{2+}$ site in the $S=2$ state. The observed value of the asymmetry parameter of EFG underlines the relevance of the tetrahedral distortion for magnetism in \lfs. The observed static magnetic hyperfine field of 14.8(3)~T indicates significant orbital contributions to the hyperfine field. Notably, both the magnetic order parameter determined by neutron diffraction and by Mössbauer spectroscopy, i.e., $\beta^{\rm PND}=0.185(10)$ and $\beta^{\rm MB}=0.116(12)$ are much smaller than expected for three-dimensional magnetic systems where $\beta \simeq 0.33$ (3D Ising) and $\beta \simeq 0.35 - 0.37$ (3D Heisenberg, 3D XY) are found~\cite{Pelissetto2002}. We emphasize, however, that $\beta^{\rm PND}$ only indirectly reflects the critical behaviour as it is affected by the decrease of the magnetic volume fraction $f_{\rm mag}$ which, in the temperature regime between 13 and 17~K, continuously decreases from 0.9 to 0.2 so that $I_{\rm B}$ is suppressed by a factor $\sqrt{f_{\rm mag}}$ (see Fig.~\ref{OP} and the detailed discussion in §~\ref{sec:MB}). The critical exponent $\beta^{\rm MB}=0.116(12)$ observed by Mössbauer spectroscopy is similar to $\beta=0.125$ of the 2D Ising model and suggests the two-dimensional nature of magnetism in \lfs.

Since orbital degrees of freedom are relevant in JT-active high-spin Fe$^{2+}$ in tetrahedral coordination with the electronic configuration $e_{\rm g}^3t_{\rm 2g}^3$, one may speculate whether a distinct orbital arrangement and/or orbital order is associated with the observed 2D nature of magnetism. The role of orbital magnetism is, e.g., relevant in the 2D Ising-type antiferromagnet FePS$_3$, where it leads to long-range antiferromagnetic order down to the monolayer limit.~\cite{Wang2016} As shown by the well-known examples KCuF$_3$ and LaMnO$_3$, particular orbital-ordered ground states are interconnected with specific (low-dimensional) magnetic structures.~\cite{Caciuffo2002,Feiner,Geck2005} This can even yield one-dimensional magnetic substructures in structurally layered systems, as seen in honeycomb-structured A$_3$Cu$_2$SbO$_6$ (A = Li, Na), where 1D magnetic substructures are formed due to the particular orbital arrangement.~\cite{Schmitt2014,Koo2016}

In summary, we have solved the magnetic ground state of \lfs\ and report experimental and theoretical evidence of the quasi-2D nature of magnetism which is due to weak and partly frustrated interlayer coupling of the rectangular $S=2$ Fe$^{2+}$ magnetic lattice. Our works adds a system with high-spin tetrahedrally coordinated Fe$^{2+}$ ions to the family of quasi-2D magnetic materials where in addition to spin also the orbital degree of freedom is relevant.

\section*{Acknowledgements}

We acknowledge support by Deutsche Forschungsgemeinschaft (DFG) under Germany's Excellence Strategy EXC2181/1-390900948 (the Heidelberg STRUCTURES Excellence Cluster) and the collaborative research center 1143 at TU Dresden. Beam time was provided by the Institut Laue-Langevin (Grenoble, France) under proposal 5-31-2539.~\cite{ill}

\bibliography{Li2FeSiO4_2024_biblio}

\begin{thebibliography}{46}%
\makeatletter
\providecommand \@ifxundefined [1]{%
 \@ifx{#1\undefined}
}%
\providecommand \@ifnum [1]{%
 \ifnum #1\expandafter \@firstoftwo
 \else \expandafter \@secondoftwo
 \fi
}%
\providecommand \@ifx [1]{%
 \ifx #1\expandafter \@firstoftwo
 \else \expandafter \@secondoftwo
 \fi
}%
\providecommand \natexlab [1]{#1}%
\providecommand \enquote  [1]{``#1''}%
\providecommand \bibnamefont  [1]{#1}%
\providecommand \bibfnamefont [1]{#1}%
\providecommand \citenamefont [1]{#1}%
\providecommand \href@noop [0]{\@secondoftwo}%
\providecommand \href [0]{\begingroup \@sanitize@url \@href}%
\providecommand \@href[1]{\@@startlink{#1}\@@href}%
\providecommand \@@href[1]{\endgroup#1\@@endlink}%
\providecommand \@sanitize@url [0]{\catcode `\\12\catcode `\$12\catcode
  `\&12\catcode `\#12\catcode `\^12\catcode `\_12\catcode `\%12\relax}%
\providecommand \@@startlink[1]{}%
\providecommand \@@endlink[0]{}%
\providecommand \url  [0]{\begingroup\@sanitize@url \@url }%
\providecommand \@url [1]{\endgroup\@href {#1}{\urlprefix }}%
\providecommand \urlprefix  [0]{URL }%
\providecommand \Eprint [0]{\href }%
\providecommand \doibase [0]{https://doi.org/}%
\providecommand \selectlanguage [0]{\@gobble}%
\providecommand \bibinfo  [0]{\@secondoftwo}%
\providecommand \bibfield  [0]{\@secondoftwo}%
\providecommand \translation [1]{[#1]}%
\providecommand \BibitemOpen [0]{}%
\providecommand \bibitemStop [0]{}%
\providecommand \bibitemNoStop [0]{.\EOS\space}%
\providecommand \EOS [0]{\spacefactor3000\relax}%
\providecommand \BibitemShut  [1]{\csname bibitem#1\endcsname}%
\let\auto@bib@innerbib\@empty
\bibitem [{\citenamefont {Gong}\ and\ \citenamefont
  {Zhang}(2019)}]{Gong2019review}%
  \BibitemOpen
  \bibfield  {author} {\bibinfo {author} {\bibfnamefont {C.}~\bibnamefont
  {Gong}}\ and\ \bibinfo {author} {\bibfnamefont {X.}~\bibnamefont {Zhang}},\
  }\bibfield  {title} {\bibinfo {title} {Two-dimensional magnetic crystals and
  emergent heterostructure devices},\ }\href
  {https://doi.org/10.1126/science.aav4450} {\bibfield  {journal} {\bibinfo
  {journal} {Science}\ }\textbf {\bibinfo {volume} {363}},\ \bibinfo {pages}
  {eaav4450} (\bibinfo {year} {2019})}\BibitemShut {NoStop}%
\bibitem [{\citenamefont {Imada}\ \emph {et~al.}(1998)\citenamefont {Imada},
  \citenamefont {Fujimori},\ and\ \citenamefont {Tokura}}]{Imada1998}%
  \BibitemOpen
  \bibfield  {author} {\bibinfo {author} {\bibfnamefont {M.}~\bibnamefont
  {Imada}}, \bibinfo {author} {\bibfnamefont {A.}~\bibnamefont {Fujimori}},\
  and\ \bibinfo {author} {\bibfnamefont {Y.}~\bibnamefont {Tokura}},\
  }\bibfield  {title} {\bibinfo {title} {Metal-insulator transitions},\ }\href
  {https://doi.org/10.1103/RevModPhys.70.1039} {\bibfield  {journal} {\bibinfo
  {journal} {Rev. Mod. Phys.}\ }\textbf {\bibinfo {volume} {70}},\ \bibinfo
  {pages} {1039} (\bibinfo {year} {1998})}\BibitemShut {NoStop}%
\bibitem [{\citenamefont {Tokura}\ and\ \citenamefont
  {Nagaosa}(2000)}]{Tokura2000}%
  \BibitemOpen
  \bibfield  {author} {\bibinfo {author} {\bibfnamefont {Y.}~\bibnamefont
  {Tokura}}\ and\ \bibinfo {author} {\bibfnamefont {N.}~\bibnamefont
  {Nagaosa}},\ }\bibfield  {title} {\bibinfo {title} {Orbital physics in
  transition-metal oxides},\ }\href
  {https://doi.org/10.1126/science.288.5465.462} {\bibfield  {journal}
  {\bibinfo  {journal} {Science}\ }\textbf {\bibinfo {volume} {288}},\ \bibinfo
  {pages} {462} (\bibinfo {year} {2000})}\BibitemShut {NoStop}%
\bibitem [{\citenamefont {Oles}(2010)}]{Oles2010}%
  \BibitemOpen
  \bibfield  {author} {\bibinfo {author} {\bibfnamefont {A.~M.}\ \bibnamefont
  {Oles}},\ }\bibfield  {title} {\bibinfo {title} {Charge and orbital order in
  transition metal oxides},\ }\href@noop {} {\bibfield  {journal} {\bibinfo
  {journal} {Acta Physica Polonica}\ }\textbf {\bibinfo {volume} {118}},\
  \bibinfo {pages} {212} (\bibinfo {year} {2010})}\BibitemShut {NoStop}%
\bibitem [{\citenamefont {Caciuffo}\ \emph {et~al.}(2002)\citenamefont
  {Caciuffo}, \citenamefont {Paolasini}, \citenamefont {Sollier}, \citenamefont
  {Ghigna}, \citenamefont {Pavarini}, \citenamefont {van~den Brink},\ and\
  \citenamefont {Altarelli}}]{Caciuffo2002}%
  \BibitemOpen
  \bibfield  {author} {\bibinfo {author} {\bibfnamefont {R.}~\bibnamefont
  {Caciuffo}}, \bibinfo {author} {\bibfnamefont {L.}~\bibnamefont {Paolasini}},
  \bibinfo {author} {\bibfnamefont {A.}~\bibnamefont {Sollier}}, \bibinfo
  {author} {\bibfnamefont {P.}~\bibnamefont {Ghigna}}, \bibinfo {author}
  {\bibfnamefont {E.}~\bibnamefont {Pavarini}}, \bibinfo {author}
  {\bibfnamefont {J.}~\bibnamefont {van~den Brink}},\ and\ \bibinfo {author}
  {\bibfnamefont {M.}~\bibnamefont {Altarelli}},\ }\bibfield  {title} {\bibinfo
  {title} {Resonant x-ray scattering study of magnetic and orbital order in
  ${\mathrm{kcuf}}_{3}$},\ }\href {https://doi.org/10.1103/PhysRevB.65.174425}
  {\bibfield  {journal} {\bibinfo  {journal} {Phys. Rev. B}\ }\textbf {\bibinfo
  {volume} {65}},\ \bibinfo {pages} {174425} (\bibinfo {year}
  {2002})}\BibitemShut {NoStop}%
\bibitem [{\citenamefont {Feiner}\ and\ \citenamefont
  {Ole\ifmmode~\acute{s}\else \'{s}\fi{}}(1999)}]{Feiner}%
  \BibitemOpen
  \bibfield  {author} {\bibinfo {author} {\bibfnamefont {L.~F.}\ \bibnamefont
  {Feiner}}\ and\ \bibinfo {author} {\bibfnamefont {A.~M.}\ \bibnamefont
  {Ole\ifmmode~\acute{s}\else \'{s}\fi{}}},\ }\bibfield  {title} {\bibinfo
  {title} {Electronic origin of magnetic and orbital ordering in insulating
  ${\mathrm{lamno}}_{3}$},\ }\href {https://doi.org/10.1103/PhysRevB.59.3295}
  {\bibfield  {journal} {\bibinfo  {journal} {Phys. Rev. B}\ }\textbf {\bibinfo
  {volume} {59}},\ \bibinfo {pages} {3295} (\bibinfo {year}
  {1999})}\BibitemShut {NoStop}%
\bibitem [{\citenamefont {Geck}\ \emph {et~al.}(2005)\citenamefont {Geck},
  \citenamefont {Wochner}, \citenamefont {Kiele}, \citenamefont {Klingeler},
  \citenamefont {Reutler}, \citenamefont {Revcolevschi},\ and\ \citenamefont
  {B\"uchner}}]{Geck2005}%
  \BibitemOpen
  \bibfield  {author} {\bibinfo {author} {\bibfnamefont {J.}~\bibnamefont
  {Geck}}, \bibinfo {author} {\bibfnamefont {P.}~\bibnamefont {Wochner}},
  \bibinfo {author} {\bibfnamefont {S.}~\bibnamefont {Kiele}}, \bibinfo
  {author} {\bibfnamefont {R.}~\bibnamefont {Klingeler}}, \bibinfo {author}
  {\bibfnamefont {P.}~\bibnamefont {Reutler}}, \bibinfo {author} {\bibfnamefont
  {A.}~\bibnamefont {Revcolevschi}},\ and\ \bibinfo {author} {\bibfnamefont
  {B.}~\bibnamefont {B\"uchner}},\ }\bibfield  {title} {\bibinfo {title}
  {Orbital polaron lattice formation in lightly doped
  ${\mathrm{la}}_{1\ensuremath{-}x}{\mathrm{sr}}_{x}{\mathrm{mno}}_{3}$},\
  }\href {https://doi.org/10.1103/PhysRevLett.95.236401} {\bibfield  {journal}
  {\bibinfo  {journal} {Phys. Rev. Lett.}\ }\textbf {\bibinfo {volume} {95}},\
  \bibinfo {pages} {236401} (\bibinfo {year} {2005})}\BibitemShut {NoStop}%
\bibitem [{\citenamefont {Islam}\ \emph {et~al.}(2011)\citenamefont {Islam},
  \citenamefont {Dominko}, \citenamefont {Masquelier}, \citenamefont
  {Sirisopanaporn}, \citenamefont {Armstrong},\ and\ \citenamefont
  {Bruce}}]{Islam2011}%
  \BibitemOpen
  \bibfield  {author} {\bibinfo {author} {\bibfnamefont {M.~S.}\ \bibnamefont
  {Islam}}, \bibinfo {author} {\bibfnamefont {R.}~\bibnamefont {Dominko}},
  \bibinfo {author} {\bibfnamefont {C.}~\bibnamefont {Masquelier}}, \bibinfo
  {author} {\bibfnamefont {C.}~\bibnamefont {Sirisopanaporn}}, \bibinfo
  {author} {\bibfnamefont {A.~R.}\ \bibnamefont {Armstrong}},\ and\ \bibinfo
  {author} {\bibfnamefont {P.~G.}\ \bibnamefont {Bruce}},\ }\bibfield  {title}
  {\bibinfo {title} {Silicate cathodes for lithium batteries: alternatives to
  phosphates?},\ }\href {https://doi.org/10.1039/C1JM10312A} {\bibfield
  {journal} {\bibinfo  {journal} {J. Mater. Chem.}\ }\textbf {\bibinfo {volume}
  {21}},\ \bibinfo {pages} {9811} (\bibinfo {year} {2011})}\BibitemShut
  {NoStop}%
\bibitem [{\citenamefont {Sirisopanaporn}\ \emph {et~al.}(2011)\citenamefont
  {Sirisopanaporn}, \citenamefont {Masquelier}, \citenamefont {Bruce},
  \citenamefont {Armstrong},\ and\ \citenamefont
  {Dominko}}]{Sirisopanaporn2011}%
  \BibitemOpen
  \bibfield  {author} {\bibinfo {author} {\bibfnamefont {C.}~\bibnamefont
  {Sirisopanaporn}}, \bibinfo {author} {\bibfnamefont {C.}~\bibnamefont
  {Masquelier}}, \bibinfo {author} {\bibfnamefont {P.~G.}\ \bibnamefont
  {Bruce}}, \bibinfo {author} {\bibfnamefont {A.~R.}\ \bibnamefont
  {Armstrong}},\ and\ \bibinfo {author} {\bibfnamefont {R.}~\bibnamefont
  {Dominko}},\ }\bibfield  {title} {\bibinfo {title} {Dependence of li2fesio4
  electrochemistry on structure},\ }\href {https://doi.org/10.1021/ja109695r}
  {\bibfield  {journal} {\bibinfo  {journal} {Journal of the American Chemical
  Society}\ }\textbf {\bibinfo {volume} {133}},\ \bibinfo {pages} {1263}
  (\bibinfo {year} {2011})},\ \bibinfo {note} {pMID: 21192677}\BibitemShut
  {NoStop}%
\bibitem [{\citenamefont {Yang}\ \emph {et~al.}(2016)\citenamefont {Yang},
  \citenamefont {Zheng}, \citenamefont {Kang}, \citenamefont {Teng},
  \citenamefont {Hu}, \citenamefont {Tan}, \citenamefont {Wang}, \citenamefont
  {Song}, \citenamefont {Xu}, \citenamefont {Mu},\ and\ \citenamefont
  {Pan}}]{Yang2016}%
  \BibitemOpen
  \bibfield  {author} {\bibinfo {author} {\bibfnamefont {J.}~\bibnamefont
  {Yang}}, \bibinfo {author} {\bibfnamefont {J.}~\bibnamefont {Zheng}},
  \bibinfo {author} {\bibfnamefont {X.}~\bibnamefont {Kang}}, \bibinfo {author}
  {\bibfnamefont {G.}~\bibnamefont {Teng}}, \bibinfo {author} {\bibfnamefont
  {L.}~\bibnamefont {Hu}}, \bibinfo {author} {\bibfnamefont {R.}~\bibnamefont
  {Tan}}, \bibinfo {author} {\bibfnamefont {K.}~\bibnamefont {Wang}}, \bibinfo
  {author} {\bibfnamefont {X.}~\bibnamefont {Song}}, \bibinfo {author}
  {\bibfnamefont {M.}~\bibnamefont {Xu}}, \bibinfo {author} {\bibfnamefont
  {S.}~\bibnamefont {Mu}},\ and\ \bibinfo {author} {\bibfnamefont
  {F.}~\bibnamefont {Pan}},\ }\bibfield  {title} {\bibinfo {title} {Tuning
  structural stability and lithium-storage properties by d-orbital
  hybridization substitution in full tetrahedron li2fesio4 nanocrystal},\
  }\href {https://doi.org/https://doi.org/10.1016/j.nanoen.2015.12.004}
  {\bibfield  {journal} {\bibinfo  {journal} {Nano Energy}\ }\textbf {\bibinfo
  {volume} {20}},\ \bibinfo {pages} {117} (\bibinfo {year} {2016})}\BibitemShut
  {NoStop}%
\bibitem [{\citenamefont {Lu}\ \emph {et~al.}(2015)\citenamefont {Lu},
  \citenamefont {Wei}, \citenamefont {Chiu}, \citenamefont {Gauvin},
  \citenamefont {Hovington}, \citenamefont {Guerfi}, \citenamefont {Zaghib},\
  and\ \citenamefont {Demopoulos}}]{Lu2015}%
  \BibitemOpen
  \bibfield  {author} {\bibinfo {author} {\bibfnamefont {X.}~\bibnamefont
  {Lu}}, \bibinfo {author} {\bibfnamefont {H.}~\bibnamefont {Wei}}, \bibinfo
  {author} {\bibfnamefont {H.-c.}\ \bibnamefont {Chiu}}, \bibinfo {author}
  {\bibfnamefont {R.}~\bibnamefont {Gauvin}}, \bibinfo {author} {\bibfnamefont
  {P.}~\bibnamefont {Hovington}}, \bibinfo {author} {\bibfnamefont
  {A.}~\bibnamefont {Guerfi}}, \bibinfo {author} {\bibfnamefont
  {K.}~\bibnamefont {Zaghib}},\ and\ \bibinfo {author} {\bibfnamefont
  {G.}~\bibnamefont {Demopoulos}},\ }\bibfield  {title} {\bibinfo {title}
  {Rate-dependent phase transitions in li2fesio4 cathode nanocrystals},\ }\href
  {https://doi.org/10.1038/srep08599} {\bibfield  {journal} {\bibinfo
  {journal} {Scientific Reports}\ }\textbf {\bibinfo {volume} {5}},\ \bibinfo
  {pages} {8599} (\bibinfo {year} {2015})}\BibitemShut {NoStop}%
\bibitem [{\citenamefont {Nishimura}\ \emph {et~al.}(2008)\citenamefont
  {Nishimura}, \citenamefont {Hayase}, \citenamefont {Kanno}, \citenamefont
  {Yashima}, \citenamefont {Nakayama},\ and\ \citenamefont
  {Yamada}}]{Nishimura2008}%
  \BibitemOpen
  \bibfield  {author} {\bibinfo {author} {\bibfnamefont {S.-i.}\ \bibnamefont
  {Nishimura}}, \bibinfo {author} {\bibfnamefont {S.}~\bibnamefont {Hayase}},
  \bibinfo {author} {\bibfnamefont {R.}~\bibnamefont {Kanno}}, \bibinfo
  {author} {\bibfnamefont {M.}~\bibnamefont {Yashima}}, \bibinfo {author}
  {\bibfnamefont {N.}~\bibnamefont {Nakayama}},\ and\ \bibinfo {author}
  {\bibfnamefont {A.}~\bibnamefont {Yamada}},\ }\bibfield  {title} {\bibinfo
  {title} {Structure of li2fesio4},\ }\href {https://doi.org/10.1021/ja805543p}
  {\bibfield  {journal} {\bibinfo  {journal} {Journal of the American Chemical
  Society}\ }\textbf {\bibinfo {volume} {130}},\ \bibinfo {pages} {13212}
  (\bibinfo {year} {2008})},\ \bibinfo {note} {pMID: 18788804}\BibitemShut
  {NoStop}%
\bibitem [{\citenamefont {Boulineau}\ \emph {et~al.}(2010)\citenamefont
  {Boulineau}, \citenamefont {Sirisopanaporn}, \citenamefont {Dominko},
  \citenamefont {Armstrong}, \citenamefont {Bruce},\ and\ \citenamefont
  {Masquelier}}]{Boulineau2010}%
  \BibitemOpen
  \bibfield  {author} {\bibinfo {author} {\bibfnamefont {A.}~\bibnamefont
  {Boulineau}}, \bibinfo {author} {\bibfnamefont {C.}~\bibnamefont
  {Sirisopanaporn}}, \bibinfo {author} {\bibfnamefont {R.}~\bibnamefont
  {Dominko}}, \bibinfo {author} {\bibfnamefont {A.~R.}\ \bibnamefont
  {Armstrong}}, \bibinfo {author} {\bibfnamefont {P.~G.}\ \bibnamefont
  {Bruce}},\ and\ \bibinfo {author} {\bibfnamefont {C.}~\bibnamefont
  {Masquelier}},\ }\bibfield  {title} {\bibinfo {title} {Polymorphism and
  structural defects in li2fesio4},\ }\href {https://doi.org/10.1039/C002815K}
  {\bibfield  {journal} {\bibinfo  {journal} {Dalton Trans.}\ }\textbf
  {\bibinfo {volume} {39}},\ \bibinfo {pages} {6310} (\bibinfo {year}
  {2010})}\BibitemShut {NoStop}%
\bibitem [{\citenamefont {Hergett}\ \emph
  {et~al.}(2019{\natexlab{a}})\citenamefont {Hergett}, \citenamefont {Jonak},
  \citenamefont {Werner}, \citenamefont {Billert}, \citenamefont {Sauerland},
  \citenamefont {Koo}, \citenamefont {Neef},\ and\ \citenamefont
  {Klingeler}}]{Hergett2019a}%
  \BibitemOpen
  \bibfield  {author} {\bibinfo {author} {\bibfnamefont {W.}~\bibnamefont
  {Hergett}}, \bibinfo {author} {\bibfnamefont {M.}~\bibnamefont {Jonak}},
  \bibinfo {author} {\bibfnamefont {J.}~\bibnamefont {Werner}}, \bibinfo
  {author} {\bibfnamefont {F.}~\bibnamefont {Billert}}, \bibinfo {author}
  {\bibfnamefont {S.}~\bibnamefont {Sauerland}}, \bibinfo {author}
  {\bibfnamefont {C.}~\bibnamefont {Koo}}, \bibinfo {author} {\bibfnamefont
  {C.}~\bibnamefont {Neef}},\ and\ \bibinfo {author} {\bibfnamefont
  {R.}~\bibnamefont {Klingeler}},\ }\bibfield  {title} {\bibinfo {title}
  {Synthesis and magnetism of a li$_2$fesio$_4$ single crystal},\ }\href
  {https://doi.org/https://doi.org/10.1016/j.jmmm.2019.01.032} {\bibfield
  {journal} {\bibinfo  {journal} {Journal of Magnetism and Magnetic Materials}\
  }\textbf {\bibinfo {volume} {477}},\ \bibinfo {pages} {1} (\bibinfo {year}
  {2019}{\natexlab{a}})}\BibitemShut {NoStop}%
\bibitem [{\citenamefont {Hergett}\ \emph
  {et~al.}(2019{\natexlab{b}})\citenamefont {Hergett}, \citenamefont {Neef},
  \citenamefont {Wadepohl}, \citenamefont {Meyer}, \citenamefont
  {Abdel-Hafiez}, \citenamefont {Ritter}, \citenamefont {Thauer},\ and\
  \citenamefont {Klingeler}}]{Hergett2019b}%
  \BibitemOpen
  \bibfield  {author} {\bibinfo {author} {\bibfnamefont {W.}~\bibnamefont
  {Hergett}}, \bibinfo {author} {\bibfnamefont {C.}~\bibnamefont {Neef}},
  \bibinfo {author} {\bibfnamefont {H.}~\bibnamefont {Wadepohl}}, \bibinfo
  {author} {\bibfnamefont {H.-P.}\ \bibnamefont {Meyer}}, \bibinfo {author}
  {\bibfnamefont {M.~M.}\ \bibnamefont {Abdel-Hafiez}}, \bibinfo {author}
  {\bibfnamefont {C.}~\bibnamefont {Ritter}}, \bibinfo {author} {\bibfnamefont
  {E.}~\bibnamefont {Thauer}},\ and\ \bibinfo {author} {\bibfnamefont
  {R.}~\bibnamefont {Klingeler}},\ }\bibfield  {title} {\bibinfo {title}
  {High-pressure optical floating-zone growth of li$_2$fesio$_4$ single
  crystals},\ }\href
  {https://doi.org/https://doi.org/10.1016/j.jcrysgro.2019.03.010} {\bibfield
  {journal} {\bibinfo  {journal} {Journal of Crystal Growth}\ }\textbf
  {\bibinfo {volume} {515}},\ \bibinfo {pages} {37} (\bibinfo {year}
  {2019}{\natexlab{b}})}\BibitemShut {NoStop}%
\bibitem [{\citenamefont {Hergett}\ \emph {et~al.}(2021)\citenamefont
  {Hergett}, \citenamefont {Neef}, \citenamefont {Meyer},\ and\ \citenamefont
  {Klingeler}}]{Hergett2021}%
  \BibitemOpen
  \bibfield  {author} {\bibinfo {author} {\bibfnamefont {W.}~\bibnamefont
  {Hergett}}, \bibinfo {author} {\bibfnamefont {C.}~\bibnamefont {Neef}},
  \bibinfo {author} {\bibfnamefont {H.-P.}\ \bibnamefont {Meyer}},\ and\
  \bibinfo {author} {\bibfnamefont {R.}~\bibnamefont {Klingeler}},\ }\bibfield
  {title} {\bibinfo {title} {Challenges in the crystal growth of
  li$_2$fesio$_4$},\ }\href
  {https://doi.org/https://doi.org/10.1016/j.jcrysgro.2020.125995} {\bibfield
  {journal} {\bibinfo  {journal} {Journal of Crystal Growth}\ }\textbf
  {\bibinfo {volume} {556}},\ \bibinfo {pages} {125995} (\bibinfo {year}
  {2021})}\BibitemShut {NoStop}%
\bibitem [{ill()}]{ill}%
  \BibitemOpen
  \href@noop {} {}\bibinfo {note} {W. Hergett, S.~Spachmann, C.~Neef,
  M.~Enderle, C.~Ritter, R.~Klingeler, Crystal and magnetic structure of
  $Pmnb$-Li$_2$FeSiO$_4$. Institut Laue-Langevin (ILL).
  doi:10.5291/ILL-DATA.5-31-2539}\BibitemShut {NoStop}%
\bibitem [{\citenamefont {Rodríguez-Carvajal}(1993)}]{Fullprof1993}%
  \BibitemOpen
  \bibfield  {author} {\bibinfo {author} {\bibfnamefont {J.}~\bibnamefont
  {Rodríguez-Carvajal}},\ }\bibfield  {title} {\bibinfo {title} {Recent
  advances in magnetic structure determination by neutron powder diffraction},\
  }\href {https://doi.org/https://doi.org/10.1016/0921-4526(93)90108-I}
  {\bibfield  {journal} {\bibinfo  {journal} {Physica B: Condensed Matter}\
  }\textbf {\bibinfo {volume} {192}},\ \bibinfo {pages} {55} (\bibinfo {year}
  {1993})}\BibitemShut {NoStop}%
\bibitem [{Bas()}]{Basireps}%
  \BibitemOpen
  \href@noop {} {}\bibinfo {note} {J. Rodriguez-Carvajal, BASIREPS: a program
  for calculating irreducible representations of space groups and basis
  functions for axial and polar vector properties. Part of the FULLPROF suite
  of programs, www.ill.eu/sites/fullprof}\BibitemShut {NoStop}%
\bibitem [{\citenamefont {Momma}\ and\ \citenamefont {Izumi}(2011)}]{Vesta}%
  \BibitemOpen
  \bibfield  {author} {\bibinfo {author} {\bibfnamefont {K.}~\bibnamefont
  {Momma}}\ and\ \bibinfo {author} {\bibfnamefont {F.}~\bibnamefont {Izumi}},\
  }\bibfield  {title} {\bibinfo {title} {{{\it VESTA3} for three-dimensional
  visualization of crystal, volumetric and morphology data}},\ }\href
  {https://doi.org/10.1107/S0021889811038970} {\bibfield  {journal} {\bibinfo
  {journal} {Journal of Applied Crystallography}\ }\textbf {\bibinfo {volume}
  {44}},\ \bibinfo {pages} {1272} (\bibinfo {year} {2011})}\BibitemShut
  {NoStop}%
\bibitem [{\citenamefont {{Kamusella}}\ and\ \citenamefont
  {{Klauss}}(2016)}]{HyInt}%
  \BibitemOpen
  \bibfield  {author} {\bibinfo {author} {\bibfnamefont {S.}~\bibnamefont
  {{Kamusella}}}\ and\ \bibinfo {author} {\bibfnamefont {H.-H.}\ \bibnamefont
  {{Klauss}}},\ }\bibfield  {title} {\bibinfo {title} {{Moessfit. A free
  M{\"o}ssbauer fitting program}},\ }\href
  {https://doi.org/10.1007/s10751-016-1247-z} {\bibfield  {journal} {\bibinfo
  {journal} {Hyperfine Interactions}\ }\textbf {\bibinfo {volume} {237}},\
  \bibinfo {eid} {82} (\bibinfo {year} {2016})}\BibitemShut {NoStop}%
\bibitem [{\citenamefont {Koepernik}\ and\ \citenamefont
  {Eschrig}(1999)}]{Koepernik_Full_potential_1999}%
  \BibitemOpen
  \bibfield  {author} {\bibinfo {author} {\bibfnamefont {K.}~\bibnamefont
  {Koepernik}}\ and\ \bibinfo {author} {\bibfnamefont {H.}~\bibnamefont
  {Eschrig}},\ }\bibfield  {title} {\bibinfo {title} {Full-potential
  nonorthogonal local-orbital minimum-basis band-structure scheme},\ }\href
  {https://doi.org/10.1103/PhysRevB.59.1743} {\bibfield  {journal} {\bibinfo
  {journal} {Phys. Rev. B}\ }\textbf {\bibinfo {volume} {59}},\ \bibinfo
  {pages} {1743} (\bibinfo {year} {1999})}\BibitemShut {NoStop}%
\bibitem [{\citenamefont {Opahle}\ \emph {et~al.}(1999)\citenamefont {Opahle},
  \citenamefont {Koepernik},\ and\ \citenamefont {Eschrig}}]{Opahle1999}%
  \BibitemOpen
  \bibfield  {author} {\bibinfo {author} {\bibfnamefont {I.}~\bibnamefont
  {Opahle}}, \bibinfo {author} {\bibfnamefont {K.}~\bibnamefont {Koepernik}},\
  and\ \bibinfo {author} {\bibfnamefont {H.}~\bibnamefont {Eschrig}},\
  }\bibfield  {title} {\bibinfo {title} {Full-potential band-structure
  calculation of iron pyrite},\ }\href
  {https://doi.org/10.1103/PhysRevB.60.14035} {\bibfield  {journal} {\bibinfo
  {journal} {Phys. Rev. B}\ }\textbf {\bibinfo {volume} {60}},\ \bibinfo
  {pages} {14035} (\bibinfo {year} {1999})}\BibitemShut {NoStop}%
\bibitem [{\citenamefont {Perdew}\ and\ \citenamefont
  {Wang}(1992)}]{Perdew_accurate_1992}%
  \BibitemOpen
  \bibfield  {author} {\bibinfo {author} {\bibfnamefont {J.~P.}\ \bibnamefont
  {Perdew}}\ and\ \bibinfo {author} {\bibfnamefont {Y.}~\bibnamefont {Wang}},\
  }\bibfield  {title} {\bibinfo {title} {Accurate and simple analytic
  representation of the electron-gas correlation energy},\ }\href
  {https://doi.org/10.1103/PhysRevB.45.13244} {\bibfield  {journal} {\bibinfo
  {journal} {Phys. Rev. B}\ }\textbf {\bibinfo {volume} {45}},\ \bibinfo
  {pages} {13244} (\bibinfo {year} {1992})}\BibitemShut {NoStop}%
\bibitem [{\citenamefont {Eschrig}\ \emph {et~al.}(2003)\citenamefont
  {Eschrig}, \citenamefont {Koepernik},\ and\ \citenamefont
  {Chaplygin}}]{Eschrig_density_2003}%
  \BibitemOpen
  \bibfield  {author} {\bibinfo {author} {\bibfnamefont {H.}~\bibnamefont
  {Eschrig}}, \bibinfo {author} {\bibfnamefont {K.}~\bibnamefont {Koepernik}},\
  and\ \bibinfo {author} {\bibfnamefont {I.}~\bibnamefont {Chaplygin}},\
  }\bibfield  {title} {\bibinfo {title} {Density functional application to
  strongly correlated electron systems},\ }\href
  {https://doi.org/https://doi.org/10.1016/S0022-4596(03)00274-3} {\bibfield
  {journal} {\bibinfo  {journal} {Journal of Solid State Chemistry}\ }\textbf
  {\bibinfo {volume} {176}},\ \bibinfo {pages} {482} (\bibinfo {year}
  {2003})},\ \bibinfo {note} {special issue on The Impact of Theoretical
  Methods on Solid-State Chemistry}\BibitemShut {NoStop}%
\bibitem [{\citenamefont {Krzystek}\ \emph {et~al.}(2006)\citenamefont
  {Krzystek}, \citenamefont {Ozarowski},\ and\ \citenamefont
  {Telser}}]{Krzystek}%
  \BibitemOpen
  \bibfield  {author} {\bibinfo {author} {\bibfnamefont {J.}~\bibnamefont
  {Krzystek}}, \bibinfo {author} {\bibfnamefont {A.}~\bibnamefont
  {Ozarowski}},\ and\ \bibinfo {author} {\bibfnamefont {J.}~\bibnamefont
  {Telser}},\ }\bibfield  {title} {\bibinfo {title} {Multi-frequency,
  high-field epr as a powerful tool to accurately determine zero-field
  splitting in high-spin transition metal coordination complexes},\ }\href
  {https://doi.org/https://doi.org/10.1016/j.ccr.2006.03.016} {\bibfield
  {journal} {\bibinfo  {journal} {Coordination Chemistry Reviews}\ }\textbf
  {\bibinfo {volume} {250}},\ \bibinfo {pages} {2308} (\bibinfo {year}
  {2006})},\ \bibinfo {note} {20th International Conference on Coordination and
  Bioinorganic Chemistry}\BibitemShut {NoStop}%
\bibitem [{\citenamefont {Gibart}\ \emph {et~al.}(1969)\citenamefont {Gibart},
  \citenamefont {Dormann},\ and\ \citenamefont {Pellerin}}]{Gibart}%
  \BibitemOpen
  \bibfield  {author} {\bibinfo {author} {\bibfnamefont {P.}~\bibnamefont
  {Gibart}}, \bibinfo {author} {\bibfnamefont {J.-L.}\ \bibnamefont
  {Dormann}},\ and\ \bibinfo {author} {\bibfnamefont {Y.}~\bibnamefont
  {Pellerin}},\ }\bibfield  {title} {\bibinfo {title} {Magnetic properties of
  fecr2s4 and cocr2s4},\ }\href
  {https://doi.org/https://doi.org/10.1002/pssb.19690360120} {\bibfield
  {journal} {\bibinfo  {journal} {physica status solidi (b)}\ }\textbf
  {\bibinfo {volume} {36}},\ \bibinfo {pages} {187} (\bibinfo {year}
  {1969})}\BibitemShut {NoStop}%
\bibitem [{\citenamefont {Fritsch}\ \emph {et~al.}(2004)\citenamefont
  {Fritsch}, \citenamefont {Hemberger}, \citenamefont {B\"uttgen},
  \citenamefont {Scheidt}, \citenamefont {Krug~von Nidda}, \citenamefont
  {Loidl},\ and\ \citenamefont {Tsurkan}}]{Fritsch}%
  \BibitemOpen
  \bibfield  {author} {\bibinfo {author} {\bibfnamefont {V.}~\bibnamefont
  {Fritsch}}, \bibinfo {author} {\bibfnamefont {J.}~\bibnamefont {Hemberger}},
  \bibinfo {author} {\bibfnamefont {N.}~\bibnamefont {B\"uttgen}}, \bibinfo
  {author} {\bibfnamefont {E.-W.}\ \bibnamefont {Scheidt}}, \bibinfo {author}
  {\bibfnamefont {H.-A.}\ \bibnamefont {Krug~von Nidda}}, \bibinfo {author}
  {\bibfnamefont {A.}~\bibnamefont {Loidl}},\ and\ \bibinfo {author}
  {\bibfnamefont {V.}~\bibnamefont {Tsurkan}},\ }\bibfield  {title} {\bibinfo
  {title} {Spin and orbital frustration in
  ${\mathrm{mnsc}}_{2}{\mathrm{s}}_{4}$ and
  ${\mathrm{fesc}}_{2}{\mathrm{s}}_{4}$},\ }\href
  {https://doi.org/10.1103/PhysRevLett.92.116401} {\bibfield  {journal}
  {\bibinfo  {journal} {Phys. Rev. Lett.}\ }\textbf {\bibinfo {volume} {92}},\
  \bibinfo {pages} {116401} (\bibinfo {year} {2004})}\BibitemShut {NoStop}%
\bibitem [{\citenamefont {Gong}\ and\ \citenamefont {Yang}(2011)}]{Gong2011}%
  \BibitemOpen
  \bibfield  {author} {\bibinfo {author} {\bibfnamefont {Z.}~\bibnamefont
  {Gong}}\ and\ \bibinfo {author} {\bibfnamefont {Y.}~\bibnamefont {Yang}},\
  }\bibfield  {title} {\bibinfo {title} {Recent advances in the research of
  polyanion-type cathode materials for li-ion batteries},\ }\href
  {https://doi.org/10.1039/C0EE00713G} {\bibfield  {journal} {\bibinfo
  {journal} {Energy Environ. Sci.}\ }\textbf {\bibinfo {volume} {4}},\ \bibinfo
  {pages} {3223} (\bibinfo {year} {2011})}\BibitemShut {NoStop}%
\bibitem [{\citenamefont {Bini}\ \emph {et~al.}(2013)\citenamefont {Bini},
  \citenamefont {Ferrari}, \citenamefont {Ferrara}, \citenamefont {Mozzati},
  \citenamefont {Capsoni}, \citenamefont {Pell}, \citenamefont {Pintacuda},\
  and\ \citenamefont {Canton}}]{Bini2013}%
  \BibitemOpen
  \bibfield  {author} {\bibinfo {author} {\bibfnamefont {M.}~\bibnamefont
  {Bini}}, \bibinfo {author} {\bibfnamefont {S.}~\bibnamefont {Ferrari}},
  \bibinfo {author} {\bibfnamefont {C.}~\bibnamefont {Ferrara}}, \bibinfo
  {author} {\bibfnamefont {M.}~\bibnamefont {Mozzati}}, \bibinfo {author}
  {\bibfnamefont {D.}~\bibnamefont {Capsoni}}, \bibinfo {author} {\bibfnamefont
  {A.}~\bibnamefont {Pell}}, \bibinfo {author} {\bibfnamefont {G.}~\bibnamefont
  {Pintacuda}},\ and\ \bibinfo {author} {\bibfnamefont {P.}~\bibnamefont
  {Canton}},\ }\bibfield  {title} {\bibinfo {title} {Polymorphism and magnetic
  properties of li2msio4 (m = fe, mn) cathode materials},\ }\href
  {https://doi.org/10.1038/srep03452} {\bibfield  {journal} {\bibinfo
  {journal} {Scientific reports}\ }\textbf {\bibinfo {volume} {3}},\ \bibinfo
  {pages} {3452} (\bibinfo {year} {2013})}\BibitemShut {NoStop}%
\bibitem [{\citenamefont {Zaghib}\ \emph {et~al.}(2006)\citenamefont {Zaghib},
  \citenamefont {Salah}, \citenamefont {Ravet}, \citenamefont {Mauger},
  \citenamefont {Gendron},\ and\ \citenamefont {Julien}}]{Zaghib2006}%
  \BibitemOpen
  \bibfield  {author} {\bibinfo {author} {\bibfnamefont {K.}~\bibnamefont
  {Zaghib}}, \bibinfo {author} {\bibfnamefont {A.}~\bibnamefont {Salah}},
  \bibinfo {author} {\bibfnamefont {N.}~\bibnamefont {Ravet}}, \bibinfo
  {author} {\bibfnamefont {A.}~\bibnamefont {Mauger}}, \bibinfo {author}
  {\bibfnamefont {F.}~\bibnamefont {Gendron}},\ and\ \bibinfo {author}
  {\bibfnamefont {C.}~\bibnamefont {Julien}},\ }\bibfield  {title} {\bibinfo
  {title} {Structural, magnetic and electrochemical properties of lithium iron
  orthosilicate},\ }\href {https://doi.org/10.1016/j.jpowsour.2006.03.023}
  {\bibfield  {journal} {\bibinfo  {journal} {Journal of Power Sources}\
  }\textbf {\bibinfo {volume} {160}},\ \bibinfo {pages} {1381} (\bibinfo {year}
  {2006})}\BibitemShut {NoStop}%
\bibitem [{\citenamefont {Werner}\ \emph {et~al.}(2020)\citenamefont {Werner},
  \citenamefont {Neef}, \citenamefont {Koo}, \citenamefont {Zvyagin},
  \citenamefont {Ponomaryov},\ and\ \citenamefont {Klingeler}}]{Werner2020}%
  \BibitemOpen
  \bibfield  {author} {\bibinfo {author} {\bibfnamefont {J.}~\bibnamefont
  {Werner}}, \bibinfo {author} {\bibfnamefont {C.}~\bibnamefont {Neef}},
  \bibinfo {author} {\bibfnamefont {C.}~\bibnamefont {Koo}}, \bibinfo {author}
  {\bibfnamefont {S.}~\bibnamefont {Zvyagin}}, \bibinfo {author} {\bibfnamefont
  {A.}~\bibnamefont {Ponomaryov}},\ and\ \bibinfo {author} {\bibfnamefont
  {R.}~\bibnamefont {Klingeler}},\ }\bibfield  {title} {\bibinfo {title}
  {Antisite disorder in the battery material ${\mathrm{lifepo}}_{4}$},\ }\href
  {https://doi.org/10.1103/PhysRevMaterials.4.115403} {\bibfield  {journal}
  {\bibinfo  {journal} {Phys. Rev. Mater.}\ }\textbf {\bibinfo {volume} {4}},\
  \bibinfo {pages} {115403} (\bibinfo {year} {2020})}\BibitemShut {NoStop}%
\bibitem [{\citenamefont {Neef}\ \emph {et~al.}(2017)\citenamefont {Neef},
  \citenamefont {Wadepohl}, \citenamefont {Meyer},\ and\ \citenamefont
  {Klingeler}}]{Neef2017}%
  \BibitemOpen
  \bibfield  {author} {\bibinfo {author} {\bibfnamefont {C.}~\bibnamefont
  {Neef}}, \bibinfo {author} {\bibfnamefont {H.}~\bibnamefont {Wadepohl}},
  \bibinfo {author} {\bibfnamefont {H.-P.}\ \bibnamefont {Meyer}},\ and\
  \bibinfo {author} {\bibfnamefont {R.}~\bibnamefont {Klingeler}},\ }\bibfield
  {title} {\bibinfo {title} {High-pressure optical floating-zone growth of
  li(mn,fe)po4 single crystals},\ }\href
  {https://doi.org/https://doi.org/10.1016/j.jcrysgro.2017.01.046} {\bibfield
  {journal} {\bibinfo  {journal} {Journal of Crystal Growth}\ }\textbf
  {\bibinfo {volume} {462}},\ \bibinfo {pages} {50} (\bibinfo {year}
  {2017})}\BibitemShut {NoStop}%
\bibitem [{\citenamefont {Köbler}(2015)}]{Koebler2015}%
  \BibitemOpen
  \bibfield  {author} {\bibinfo {author} {\bibfnamefont {U.}~\bibnamefont
  {Köbler}},\ }\bibfield  {title} {\bibinfo {title} {{One-Dimensional Boson
  Fields in the Critical Range of EuS and EuO}},\ }\href
  {https://doi.org/10.12693/APhysPolA.128.398} {\bibfield  {journal} {\bibinfo
  {journal} {Acta Physica Polonica A}\ }\textbf {\bibinfo {volume} {128}},\
  \bibinfo {pages} {398} (\bibinfo {year} {2015})}\BibitemShut {NoStop}%
\bibitem [{hyp(2007)}]{hyperfineinteraction}%
  \BibitemOpen
  \bibinfo {title} {Hyperfine interactions},\ in\ \href
  {https://doi.org/10.1002/9783527611423.ch2} {\emph {\bibinfo {booktitle}
  {Mössbauer Effect in Lattice Dynamics}}}\ (\bibinfo  {publisher} {John Wiley
  \& Sons, Ltd},\ \bibinfo {year} {2007})\ Chap.~\bibinfo {chapter} {2}, pp.\
  \bibinfo {pages} {29--77}\BibitemShut {NoStop}%
\bibitem [{\citenamefont {Imbert}\ \emph {et~al.}(1973)\citenamefont {Imbert},
  \citenamefont {Varret},\ and\ \citenamefont {Wintenberger}}]{imbert}%
  \BibitemOpen
  \bibfield  {author} {\bibinfo {author} {\bibfnamefont {P.}~\bibnamefont
  {Imbert}}, \bibinfo {author} {\bibfnamefont {F.}~\bibnamefont {Varret}},\
  and\ \bibinfo {author} {\bibfnamefont {M.}~\bibnamefont {Wintenberger}},\
  }\bibfield  {title} {\bibinfo {title} {Etude par effet mössbauer de la
  briartite (cu2feges4)},\ }\href
  {https://doi.org/https://doi.org/10.1016/S0022-3697(73)80133-7} {\bibfield
  {journal} {\bibinfo  {journal} {Journal of Physics and Chemistry of Solids}\
  }\textbf {\bibinfo {volume} {34}},\ \bibinfo {pages} {1675 } (\bibinfo {year}
  {1973})}\BibitemShut {NoStop}%
\bibitem [{\citenamefont {Sa\'ul}\ and\ \citenamefont
  {Radtke}(2011)}]{Saul_magnetic_2011}%
  \BibitemOpen
  \bibfield  {author} {\bibinfo {author} {\bibfnamefont {A.}~\bibnamefont
  {Sa\'ul}}\ and\ \bibinfo {author} {\bibfnamefont {G.}~\bibnamefont
  {Radtke}},\ }\bibfield  {title} {\bibinfo {title} {Magnetic couplings in
  ${\mathrm{csv}}_{2}{\mathrm{o}}_{5}$: A new picture},\ }\href
  {https://doi.org/10.1103/PhysRevLett.106.177203} {\bibfield  {journal}
  {\bibinfo  {journal} {Phys. Rev. Lett.}\ }\textbf {\bibinfo {volume} {106}},\
  \bibinfo {pages} {177203} (\bibinfo {year} {2011})}\BibitemShut {NoStop}%
\bibitem [{\citenamefont {Sa\'ul}\ and\ \citenamefont
  {Radtke}(2014)}]{Saul2014}%
  \BibitemOpen
  \bibfield  {author} {\bibinfo {author} {\bibfnamefont {A.}~\bibnamefont
  {Sa\'ul}}\ and\ \bibinfo {author} {\bibfnamefont {G.}~\bibnamefont
  {Radtke}},\ }\bibfield  {title} {\bibinfo {title} {Density functional
  approach for the magnetism of $\ensuremath{\beta}$-tevo${}_{4}$},\ }\href
  {https://doi.org/10.1103/PhysRevB.89.104414} {\bibfield  {journal} {\bibinfo
  {journal} {Phys. Rev. B}\ }\textbf {\bibinfo {volume} {89}},\ \bibinfo
  {pages} {104414} (\bibinfo {year} {2014})}\BibitemShut {NoStop}%
\bibitem [{\citenamefont {Zvereva}\ \emph {et~al.}(2015)\citenamefont
  {Zvereva}, \citenamefont {Stratan}, \citenamefont {Ovchenkov}, \citenamefont
  {Nalbandyan}, \citenamefont {Lin}, \citenamefont {Vavilova}, \citenamefont
  {Iakovleva}, \citenamefont {Abdel-Hafiez}, \citenamefont {Silhanek},
  \citenamefont {Chen}, \citenamefont {Stroppa}, \citenamefont {Picozzi},
  \citenamefont {Jeschke}, \citenamefont {Valent\'{\i}},\ and\ \citenamefont
  {Vasiliev}}]{Zvereva_zigzag_2015}%
  \BibitemOpen
  \bibfield  {author} {\bibinfo {author} {\bibfnamefont {E.~A.}\ \bibnamefont
  {Zvereva}}, \bibinfo {author} {\bibfnamefont {M.~I.}\ \bibnamefont
  {Stratan}}, \bibinfo {author} {\bibfnamefont {Y.~A.}\ \bibnamefont
  {Ovchenkov}}, \bibinfo {author} {\bibfnamefont {V.~B.}\ \bibnamefont
  {Nalbandyan}}, \bibinfo {author} {\bibfnamefont {J.-Y.}\ \bibnamefont {Lin}},
  \bibinfo {author} {\bibfnamefont {E.~L.}\ \bibnamefont {Vavilova}}, \bibinfo
  {author} {\bibfnamefont {M.~F.}\ \bibnamefont {Iakovleva}}, \bibinfo {author}
  {\bibfnamefont {M.}~\bibnamefont {Abdel-Hafiez}}, \bibinfo {author}
  {\bibfnamefont {A.~V.}\ \bibnamefont {Silhanek}}, \bibinfo {author}
  {\bibfnamefont {X.-J.}\ \bibnamefont {Chen}}, \bibinfo {author}
  {\bibfnamefont {A.}~\bibnamefont {Stroppa}}, \bibinfo {author} {\bibfnamefont
  {S.}~\bibnamefont {Picozzi}}, \bibinfo {author} {\bibfnamefont {H.~O.}\
  \bibnamefont {Jeschke}}, \bibinfo {author} {\bibfnamefont {R.}~\bibnamefont
  {Valent\'{\i}}},\ and\ \bibinfo {author} {\bibfnamefont {A.~N.}\ \bibnamefont
  {Vasiliev}},\ }\bibfield  {title} {\bibinfo {title} {Zigzag antiferromagnetic
  quantum ground state in monoclinic honeycomb lattice antimonates
  ${A}_{3}\mathrm{N}{\mathrm{i}}_{2}\mathrm{Sb}{\mathrm{o}}_{6}\phantom{\rule{0.28em}{0ex}}(a=\mathrm{Li},\phantom{\rule{0.28em}{0ex}}\mathrm{Na})$},\
  }\href {https://doi.org/10.1103/PhysRevB.92.144401} {\bibfield  {journal}
  {\bibinfo  {journal} {Phys. Rev. B}\ }\textbf {\bibinfo {volume} {92}},\
  \bibinfo {pages} {144401} (\bibinfo {year} {2015})}\BibitemShut {NoStop}%
\bibitem [{\citenamefont {Foyevtsova}\ \emph {et~al.}(2011)\citenamefont
  {Foyevtsova}, \citenamefont {Opahle}, \citenamefont {Zhang}, \citenamefont
  {Jeschke},\ and\ \citenamefont
  {Valent\'{\i}}}]{Foyevtsova_determination_2011}%
  \BibitemOpen
  \bibfield  {author} {\bibinfo {author} {\bibfnamefont {K.}~\bibnamefont
  {Foyevtsova}}, \bibinfo {author} {\bibfnamefont {I.}~\bibnamefont {Opahle}},
  \bibinfo {author} {\bibfnamefont {Y.-Z.}\ \bibnamefont {Zhang}}, \bibinfo
  {author} {\bibfnamefont {H.~O.}\ \bibnamefont {Jeschke}},\ and\ \bibinfo
  {author} {\bibfnamefont {R.}~\bibnamefont {Valent\'{\i}}},\ }\bibfield
  {title} {\bibinfo {title} {Determination of effective microscopic models for
  the frustrated antiferromagnets cs${}_{2}$cucl${}_{4}$ and
  cs${}_{2}$cubr${}_{4}$ by density functional methods},\ }\href
  {https://doi.org/10.1103/PhysRevB.83.125126} {\bibfield  {journal} {\bibinfo
  {journal} {Phys. Rev. B}\ }\textbf {\bibinfo {volume} {83}},\ \bibinfo
  {pages} {125126} (\bibinfo {year} {2011})}\BibitemShut {NoStop}%
\bibitem [{\citenamefont {Inc.}()}]{Mathematica}%
  \BibitemOpen
  \bibfield  {author} {\bibinfo {author} {\bibfnamefont {W.~R.}\ \bibnamefont
  {Inc.}},\ }\href@noop {} {\bibinfo {title} {Mathematica, {V}ersion 11.3}},\
  \bibinfo {note} {champaign, IL, 2018}\BibitemShut {NoStop}%
\bibitem [{\citenamefont {de~Jongh}\ and\ \citenamefont
  {Miedema}(1974)}]{dejongh1974}%
  \BibitemOpen
  \bibfield  {author} {\bibinfo {author} {\bibfnamefont {L.}~\bibnamefont
  {de~Jongh}}\ and\ \bibinfo {author} {\bibfnamefont {A.}~\bibnamefont
  {Miedema}},\ }\bibfield  {title} {\bibinfo {title} {Experiments on simple
  magnetic model systems},\ }\href {https://doi.org/10.1080/00018739700101558}
  {\bibfield  {journal} {\bibinfo  {journal} {Advances in Physics}\ }\textbf
  {\bibinfo {volume} {23}},\ \bibinfo {pages} {1} (\bibinfo {year}
  {1974})}\BibitemShut {NoStop}%
\bibitem [{\citenamefont {Pelissetto}\ and\ \citenamefont
  {Vicari}(2002)}]{Pelissetto2002}%
  \BibitemOpen
  \bibfield  {author} {\bibinfo {author} {\bibfnamefont {A.}~\bibnamefont
  {Pelissetto}}\ and\ \bibinfo {author} {\bibfnamefont {E.}~\bibnamefont
  {Vicari}},\ }\bibfield  {title} {\bibinfo {title} {Critical phenomena and
  renormalization-group theory},\ }\href
  {https://doi.org/https://doi.org/10.1016/S0370-1573(02)00219-3} {\bibfield
  {journal} {\bibinfo  {journal} {Physics Reports}\ }\textbf {\bibinfo {volume}
  {368}},\ \bibinfo {pages} {549} (\bibinfo {year} {2002})}\BibitemShut
  {NoStop}%
\bibitem [{\citenamefont {Wang}\ \emph {et~al.}(2016)\citenamefont {Wang},
  \citenamefont {Du}, \citenamefont {Liu}, \citenamefont {Hu}, \citenamefont
  {Zhang}, \citenamefont {Zhang}, \citenamefont {Owen}, \citenamefont {Lu},
  \citenamefont {Gan}, \citenamefont {Sengupta}, \citenamefont {Kloc},\ and\
  \citenamefont {Xiong}}]{Wang2016}%
  \BibitemOpen
  \bibfield  {author} {\bibinfo {author} {\bibfnamefont {X.}~\bibnamefont
  {Wang}}, \bibinfo {author} {\bibfnamefont {K.}~\bibnamefont {Du}}, \bibinfo
  {author} {\bibfnamefont {F.}~\bibnamefont {Liu}}, \bibinfo {author}
  {\bibfnamefont {P.}~\bibnamefont {Hu}}, \bibinfo {author} {\bibfnamefont
  {J.}~\bibnamefont {Zhang}}, \bibinfo {author} {\bibfnamefont
  {Q.}~\bibnamefont {Zhang}}, \bibinfo {author} {\bibfnamefont
  {M.}~\bibnamefont {Owen}}, \bibinfo {author} {\bibfnamefont {X.}~\bibnamefont
  {Lu}}, \bibinfo {author} {\bibfnamefont {C.}~\bibnamefont {Gan}}, \bibinfo
  {author} {\bibfnamefont {P.}~\bibnamefont {Sengupta}}, \bibinfo {author}
  {\bibfnamefont {C.}~\bibnamefont {Kloc}},\ and\ \bibinfo {author}
  {\bibfnamefont {Q.}~\bibnamefont {Xiong}},\ }\bibfield  {title} {\bibinfo
  {title} {Raman spectroscopy of atomically thin two-dimensional magnetic iron
  phosphorus trisulfide (feps3) crystals},\ }\href
  {https://doi.org/10.1088/2053-1583/3/3/031009} {\bibfield  {journal}
  {\bibinfo  {journal} {2D Materials}\ }\textbf {\bibinfo {volume} {3}},\
  \bibinfo {pages} {031009} (\bibinfo {year} {2016})}\BibitemShut {NoStop}%
\bibitem [{\citenamefont {Schmitt}\ \emph {et~al.}(2014)\citenamefont
  {Schmitt}, \citenamefont {Janson}, \citenamefont {Golbs}, \citenamefont
  {Schmidt}, \citenamefont {Schnelle}, \citenamefont {Richter},\ and\
  \citenamefont {Rosner}}]{Schmitt2014}%
  \BibitemOpen
  \bibfield  {author} {\bibinfo {author} {\bibfnamefont {M.}~\bibnamefont
  {Schmitt}}, \bibinfo {author} {\bibfnamefont {O.}~\bibnamefont {Janson}},
  \bibinfo {author} {\bibfnamefont {S.}~\bibnamefont {Golbs}}, \bibinfo
  {author} {\bibfnamefont {M.}~\bibnamefont {Schmidt}}, \bibinfo {author}
  {\bibfnamefont {W.}~\bibnamefont {Schnelle}}, \bibinfo {author}
  {\bibfnamefont {J.}~\bibnamefont {Richter}},\ and\ \bibinfo {author}
  {\bibfnamefont {H.}~\bibnamefont {Rosner}},\ }\bibfield  {title} {\bibinfo
  {title} {Microscopic magnetic modeling for the
  $s\phantom{\rule{0.16em}{0ex}}=\phantom{\rule{0.16em}{0ex}}\frac{1}{2}$
  alternating-chain compounds
  ${\mathrm{na}}_{3}{\mathrm{cu}}_{2}{\mathrm{sbo}}_{6}$ and
  na${}_{2}$cu${}_{2}$teo${}_{6}$},\ }\href
  {https://doi.org/10.1103/PhysRevB.89.174403} {\bibfield  {journal} {\bibinfo
  {journal} {Phys. Rev. B}\ }\textbf {\bibinfo {volume} {89}},\ \bibinfo
  {pages} {174403} (\bibinfo {year} {2014})}\BibitemShut {NoStop}%
\bibitem [{\citenamefont {Koo}\ \emph {et~al.}(2016)\citenamefont {Koo},
  \citenamefont {Zvereva}, \citenamefont {Shukaev}, \citenamefont {Richter},
  \citenamefont {Stratan}, \citenamefont {Vasiliev}, \citenamefont
  {Nalbandyan},\ and\ \citenamefont {Klingeler}}]{Koo2016}%
  \BibitemOpen
  \bibfield  {author} {\bibinfo {author} {\bibfnamefont {C.}~\bibnamefont
  {Koo}}, \bibinfo {author} {\bibfnamefont {E.}~\bibnamefont {Zvereva}},
  \bibinfo {author} {\bibfnamefont {I.}~\bibnamefont {Shukaev}}, \bibinfo
  {author} {\bibfnamefont {M.}~\bibnamefont {Richter}}, \bibinfo {author}
  {\bibfnamefont {M.}~\bibnamefont {Stratan}}, \bibinfo {author} {\bibfnamefont
  {A.}~\bibnamefont {Vasiliev}}, \bibinfo {author} {\bibfnamefont
  {V.}~\bibnamefont {Nalbandyan}},\ and\ \bibinfo {author} {\bibfnamefont
  {R.}~\bibnamefont {Klingeler}},\ }\bibfield  {title} {\bibinfo {title}
  {Static and dynamic magnetic response of fragmented haldane-like spin chains
  in layered li3cu2sbo6},\ }\href {https://doi.org/10.7566/JPSJ.85.084702}
  {\bibfield  {journal} {\bibinfo  {journal} {Journal of the Physical Society
  of Japan}\ }\textbf {\bibinfo {volume} {85}},\ \bibinfo {pages} {084702}
  (\bibinfo {year} {2016})}\BibitemShut {NoStop}%
\end{thebibliography}%

\end{document}